\documentclass[pdflatex,sn-nature]{sn-jnl}

\usepackage{graphicx}%
\usepackage{multirow}%
\usepackage{amsmath,amssymb,amsfonts}%
\usepackage{xcolor}%
\usepackage{textcomp}%
\usepackage{manyfoot}%
\usepackage{booktabs}%

\newcommand{\panel}[3]{%
  \begin{minipage}[t]{#1}
    {\small #3}\\[1pt]
    \includegraphics[width=\linewidth]{#2}
  \end{minipage}%
}

\begin{document}

\title[FlakeForge: automated device layout for exfoliated flakes]{FlakeForge: automated lithography-ready device layout from optical images of exfoliated van der Waals flakes}

\author*[1]{\fnm{Or} \sur{Berman}}\email{orberman5@gmail.com}
\equalcont{These authors contributed equally to this work.}
\author[1]{\fnm{Ilay} \sur{Mangel}}
\equalcont{These authors contributed equally to this work.}

\affil*[1]{\orgdiv{Department of Physics}, \orgname{Technion -- Israel Institute of Technology}, \orgaddress{\city{Haifa}, \postcode{3200003  }, \country{Israel}}}

\abstract{
Fabricating a device from an exfoliated van der Waals flake requires finding a suitable flake under a microscope and designing its lithography layout by hand. We present FlakeForge, an open-source pipeline from raw optical images of a marker-patterned chip to a lithography-ready, multilayer design file. It scores flakes for device usability, aligns and scales each image from its markers, fits the user's device design to each selected flake's contour and connects every contact to a bond pad by a trace routed around all earlier devices, so no two overlap. Demonstrated on Hall bars and Little-Parks rings, it processed four chips in about half an hour, work that took an expert days. It runs fully unattended or pauses for user review. Adapting it to a new material, marker design or device dimensions requires only configuration, with no training data; a new device type requires only its geometry definition.}


\maketitle

\section{Introduction}\label{sec:intro}

Mechanical exfoliation remains the standard way to obtain the atomically thin flakes used in van der Waals (vdW) materials-based devices, because it yields clean flakes without the equipment that growth requires. It is also uncontrolled: flakes land with random size, shape, thickness and position, so no two chips are alike, nor are any two flakes. Device fabrication therefore begins with a researcher scanning the chip under an optical microscope and judging which flakes suit the planned device: clean, of uniform thickness and isolated enough to contact. Each chosen flake then needs its own fabrication layers designed by hand in mask layout software. The process is slow, varies between researchers and scales poorly when a study needs many devices, whether for statistics or for optimizing the device geometry.

Thin flakes are visible on a suitably chosen substrate through thin-film interference\cite{blake2007,hattori2024hbncolor}, and most automated identification tools detect flakes from the colour contrast this produces. These tools now process full-chip scans using unsupervised clustering\cite{sterbentz2021universalsegmentation}, Gaussian mixture models\cite{uslu2023gmm} or deep-learning segmentation\cite{masubuchi2020deeplearning,uslu2024maskterial}, supported by large labelled datasets\cite{yan2025flakedataset}. Current work extends them in several directions: reducing the per-material training they require\cite{yang2025atomic,nguyen2026qupaint,pandey2025cliff}, increasing their speed\cite{gasbarro2025highthroughput}, running them on the microscope's own computer\cite{tang2026lightweightmodel} and replacing reflected colour with photoluminescence as the detection signal\cite{crimmann2025tmdidentification}. These tools identify which regions of an image are flakes of a given thickness, and some report further properties such as size or layer number. They do not, in general, decide whether a flake suits a specific device, because the properties that matter depend on the device geometry: the flake's shape, the cleanliness of its surface and what lies around it. That judgement remains with the researcher.

Automation has also moved beyond detection. A tabletop instrument exfoliates crystals at a throughput comparable to that of an experienced researcher and locates monolayers automatically\cite{courtney2024exfoliator}, and robotic systems search for flakes and stack them into heterostructures\cite{masubuchi2018autonomous,li2026roboticstacking}. Li et al.\cite{li2025flakestacking} select usable flakes and record their coordinates, geometry and thickness to plan automated pick-up and stacking, and OpenQlaw\cite{pandey2026openqlaw} uses a language-model agent to carry identification into downstream laboratory operations. Recent reviews describe this automation as advancing step by step rather than as one system that carries a flake to a finished device\cite{davari2026roboticreview,ge2025aiexfoliation}.



None of these systems generates the fabrication layers that turn each flake into a device, such as its etch mask and metallization. The embodied artificial-intelligence system of Shi et al.~\cite{shi2026qumus} plans and executes the vdW stacking for a target device geometry. Its contact layout is supplied to it rather than generated, and it requires a dedicated robotic platform.

The closest work is that of Lee et al.~\cite{lee2026mos2layout}, who generated source--drain layouts for MoS$_2$ transistors. Their layout is a fixed two-terminal contact pattern that connects each flake to one pair of prepatterned bond pads. Because the pads are fixed in advance, the layout cannot adapt to how many terminals a device needs or to where a flake lands: flakes too close to one another are excluded. It also writes no etch layer, so each flake keeps the shape in which it landed.

Here we present FlakeForge, an open-source pipeline that takes raw optical images of exfoliated flakes, on a chip prepatterned with an alignment-marker grid, to a lithography-ready layout in the Caltech Intermediate Form (CIF) format. It works in three stages. Flake detection scores every candidate with a configurable usability measure that combines colour, surface texture, shape and isolation from neighbouring material, and selects the flakes to build on. Alignment and scaling locates the two top markers visible in each image and converts their pixel coordinates to chip coordinates. Device layout fits the user's device design to the contour of each selected flake (sorted by usability score) and connects every contact to its own bond pad while avoiding all geometry already placed on the chip. The layout can contain any number of user-defined fabrication layers. We implemented two device types, four-terminal Little-Parks (LP) rings\cite{littleparks1962} and eight-terminal Hall bars, each written as a three-layer design: two layers for the fine and coarse electron-beam lithography (EBL) exposures of the metallization, and one for the EBL exposure of the etch mask. Each stage reads only the output of the previous one and is configured separately, so it can be retuned or replaced without changes elsewhere, and the detection and alignment stages can pause for a user to accept, correct or reject their decisions. The pipeline needs no labelled data from its user and no graphics processing unit (GPU). Run unattended on four chips of exfoliated 4Hb-TaS$_2$, it produced 119 device layouts in 34 minutes on a standard laptop, and a domain expert then reviewed every layout individually.

\section{Results}\label{sec:results}

\subsection{A pipeline from chip images to a layout file}\label{sec:overview}

\begin{figure}[htb]
\centering
\includegraphics[width=\textwidth]{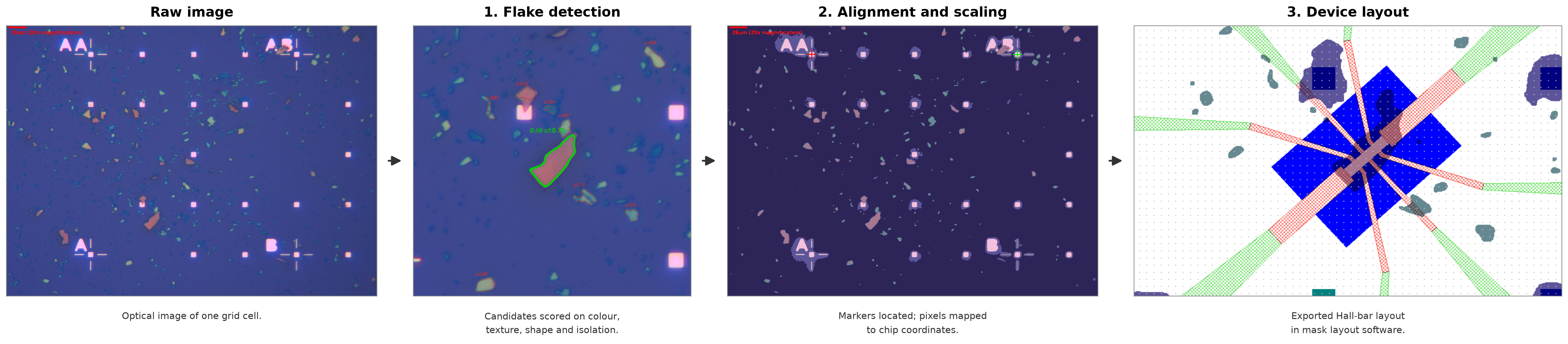}
\caption{\textbf{Overview of the FlakeForge pipeline.} 
Flake detection scores candidate flakes for usability; alignment and scaling locates the two crosshair markers in each image, converts pixels to chip coordinates and extracts flake polygons into the chip layout; device layout fits Hall-bar or Little--Parks ring geometry to each selected flake and routes its contacts to pads. The device-layout panel is exported CIF geometry viewed in mask layout software, for a Hall bar on a different flake from the first three panels (blue, etch layer; red hatching, internal contacts; green hatching, external arms).}
\label{fig:overview}
\end{figure}

FlakeForge runs in three stages (Fig.~\ref{fig:overview}): flake detection, alignment and scaling, and device layout. The input is a folder of optical images of one chip. The chip carries a square grid of alignment markers on a 200~$\mu$m pitch, with a crosshair at each cell corner and smaller markers spaced 50~$\mu$m apart inside each cell. Each image is named after the grid cell of its left-hand crosshair. This naming rule gives every image an absolute position on the chip before anything in it is detected, and gives every flake and device an address shared by the user and the program (\hyperref[sec:m-naming]{Methods}).

The interface between stages is deliberately narrow. Detection passes on a pixel contour and a score per flake; alignment and scaling passes on calibrated polygons placed in a copy of the chip's master layout; device layout writes the device geometry into that file. No stage inspects how its input was produced, so the detector, for example, can be swapped for a method designed for monolayers\cite{uslu2023gmm,uslu2024maskterial} without changes downstream.

\subsection{Selecting flakes by usability}\label{sec:detection-results}

The detector is built to decide whether a flake is worth fabricating a device on. Candidates are filtered by area and shape and then scored on three criteria (\hyperref[sec:m-detection]{Methods}). Colour fraction measures how much of the flake lies within the colour range of the target thickness. Texture combines an edge-density focus measure\cite{pechpacheco2000} with the intensity spread inside the flake, so that cracks, contamination (residue, dust or substrate fragments) and thickness steps lower the score. Isolation measures how much of a padded neighbourhood around the flake is free of other material, to decide whether contacts and traces can later be placed without shorting through neighbouring flakes or markers. A weighted sum of the three gives the final score, which has its own threshold, so a flake can pass every individual test and still be rejected.

\begin{figure}[htb]
\centering
\includegraphics[width=\textwidth]{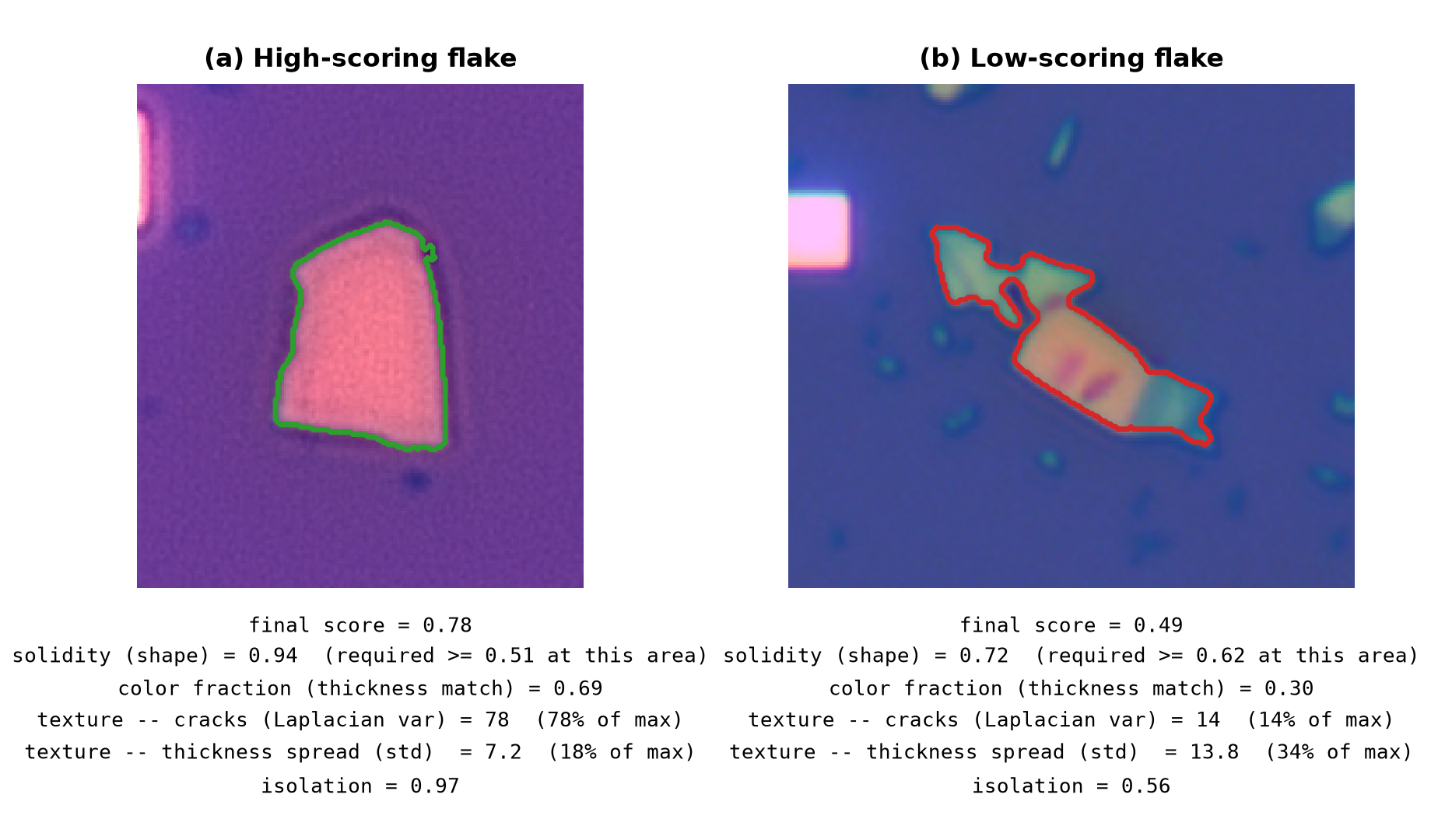}

\vspace{6pt}
\panel{0.32\textwidth}{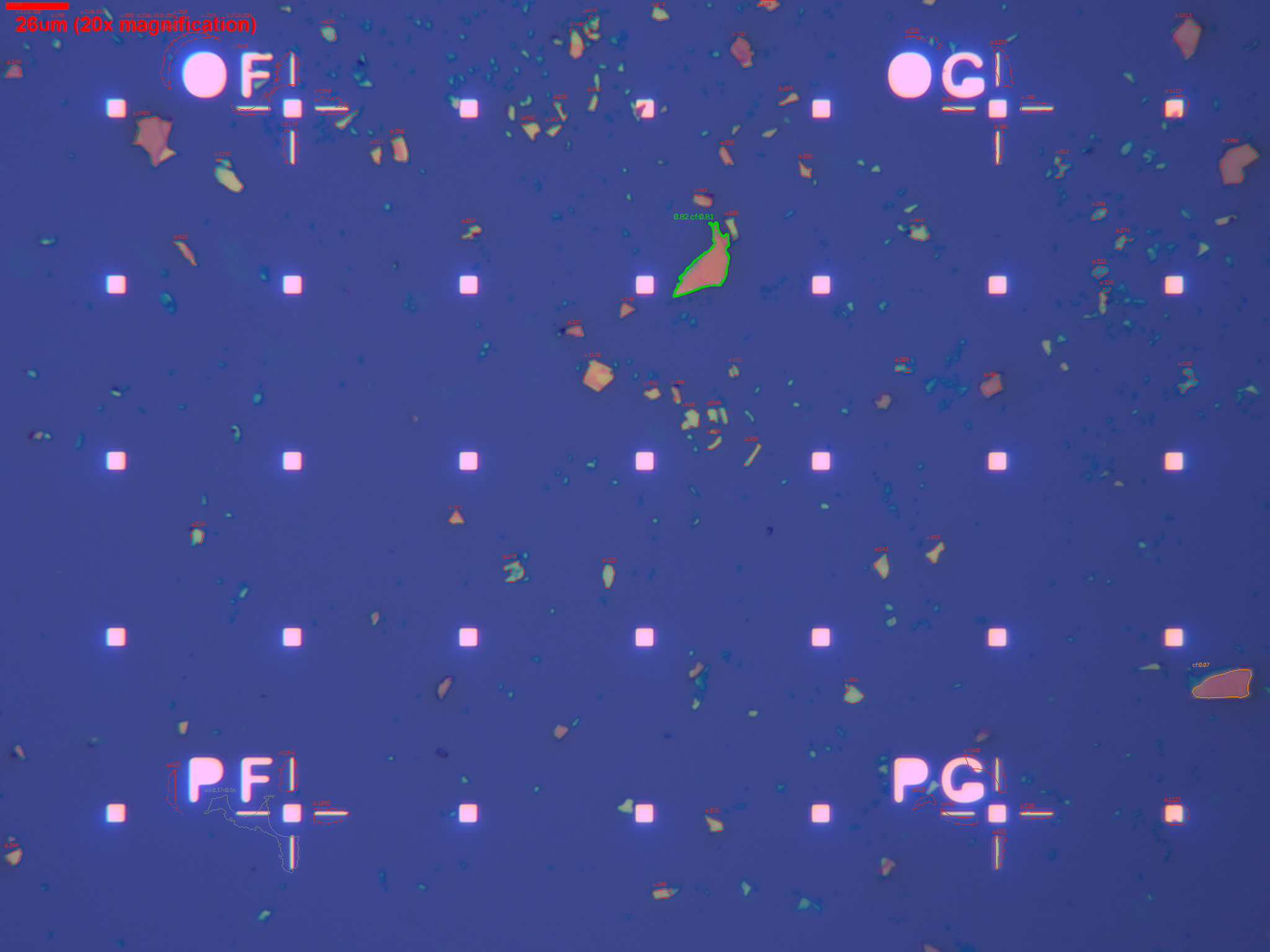}{\textbf{c.}}
\hfill
\panel{0.32\textwidth}{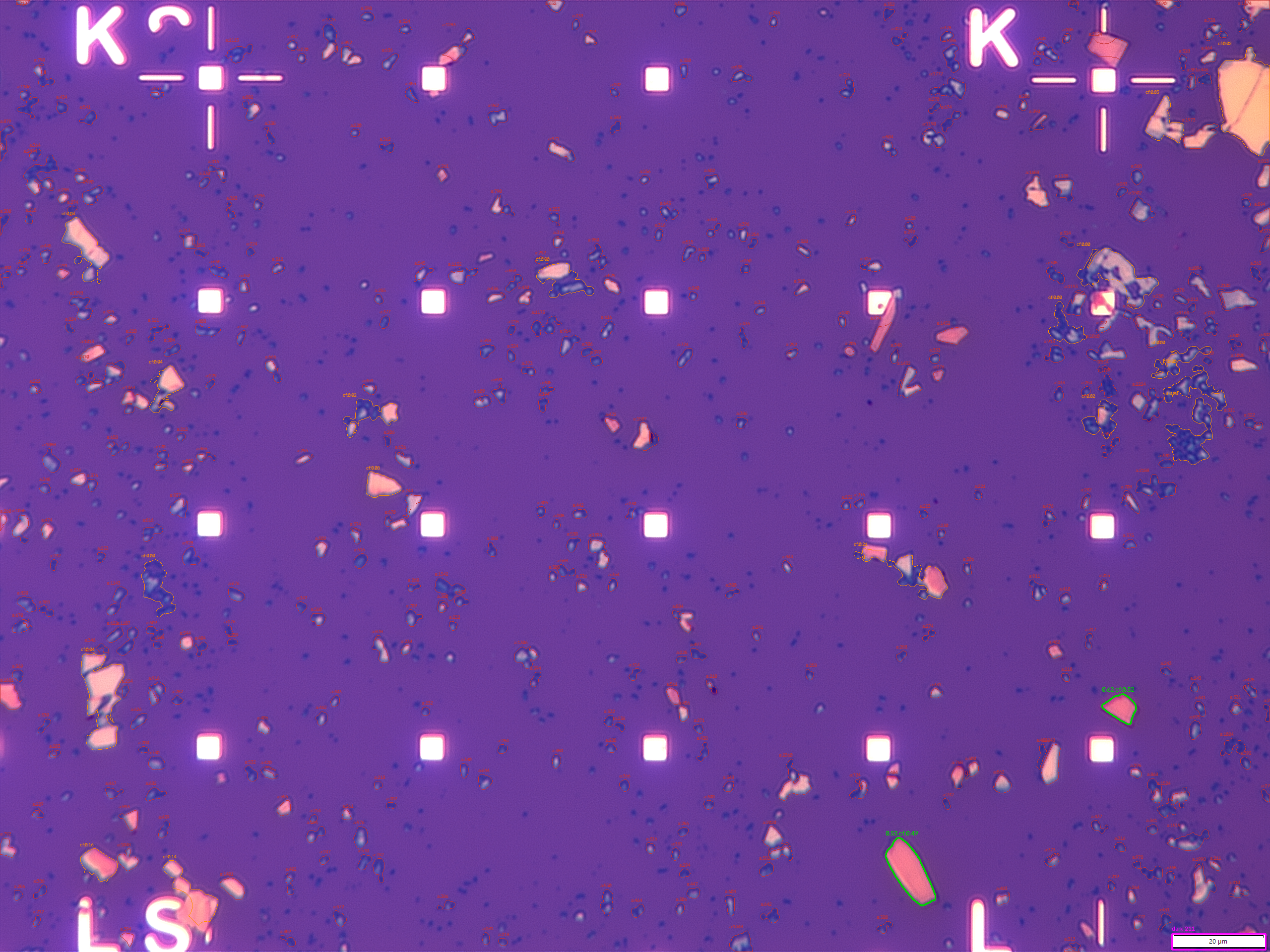}{\textbf{d.}}
\hfill
\panel{0.32\textwidth}{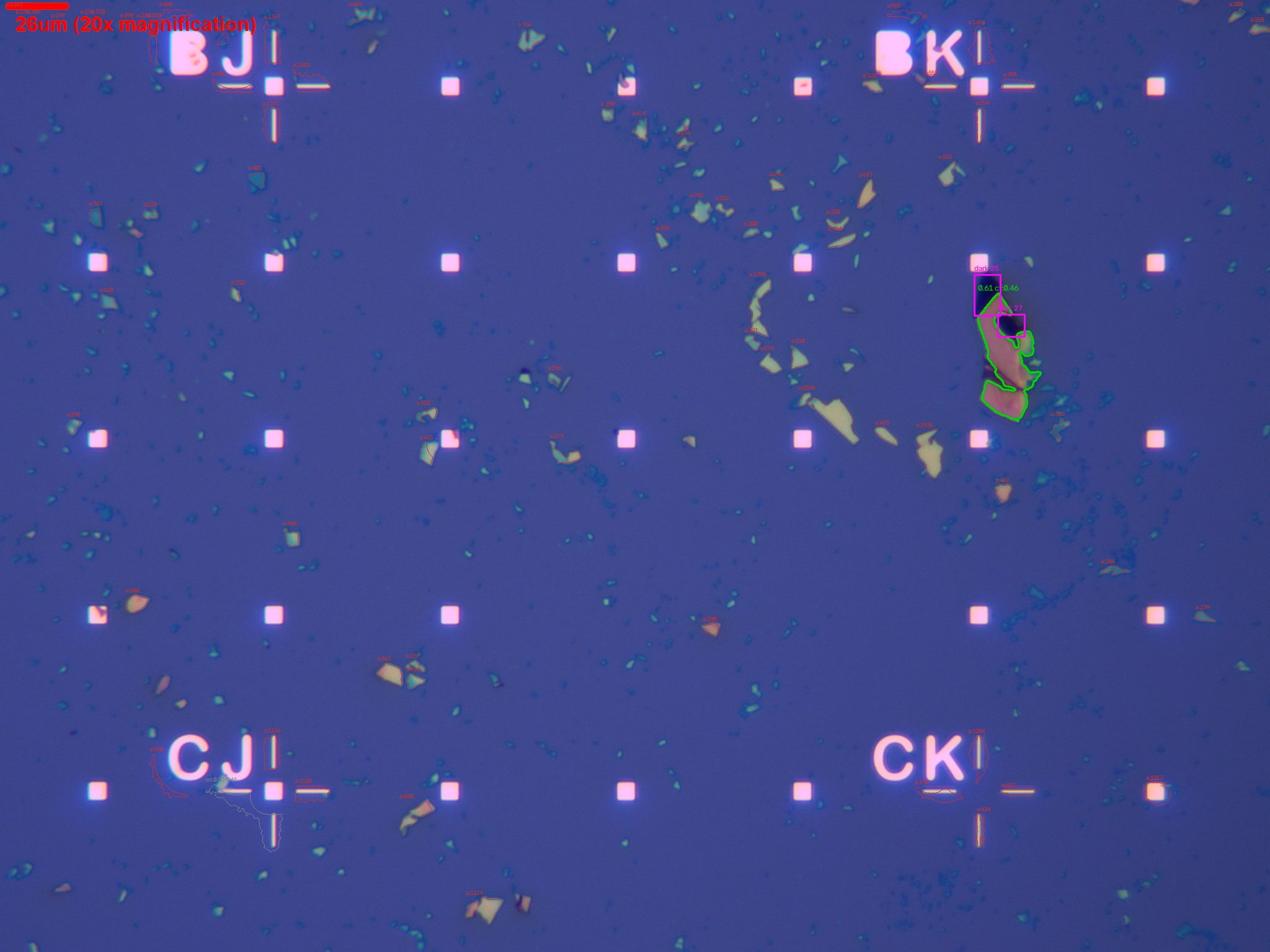}{\textbf{e.}}
\caption{\textbf{Usability scoring and its diagnostic display.} \textbf{a},~\textbf{b} Two real candidates from different chips, each scored under its own chip's configuration: \textbf{a} a clean, well-isolated flake (score 0.78); \textbf{b} a candidate scoring 0.49, below its chip's colour-fraction threshold (0.30 against 0.35) and poorly isolated (0.56 against 0.97 in \textbf{a}). Both pass the solidity test (0.94 and 0.72 against size-relaxed thresholds of 0.51 and 0.62), although the shape in \textbf{b} would make contact placement harder. Texture values are also given as a percentage of their configured maxima. \textbf{c},~\textbf{d} Contours drawn in the colour of the first test they failed (red, area; blue, touching the image border; grey, solidity; orange, colour fraction; yellow or purple, texture or isolation, whichever was weaker) or in green with a thicker outline if they passed, on two chips imaged with different camera settings. \textbf{e} Dark-region alert (magenta box) next to an accepted flake. }
\label{fig:selection}
\end{figure}

Figure~\ref{fig:selection}a,b compares two real candidates, each scored under its own chip's configuration. The higher-scoring flake (0.78) is clean and well isolated; the lower-scoring one (0.49) falls below the colour-fraction threshold and shares its neighbourhood with similarly coloured fragments (isolation 0.56 against 0.97). Area and shape filters run before scoring, including a solidity test (contour area over convex-hull area) that rejects irregular shapes on which contacts are hard to place. Its threshold relaxes as candidates grow, because a long, gently bent flake is legitimately concave.

The detector draws every evaluated contour in the colour of the first test it failed, or in green if it passed every test (Fig.~\ref{fig:selection}c,d). The user therefore sees where each candidate was lost rather than only the final decision, which makes the thresholds easier to tune. A separate alert marks very dark regions, which usually indicate thick material or fabrication defects, without affecting the selection (Fig.~\ref{fig:selection}e). Lee et al.\cite{lee2026mos2layout} use area, solidity and intensity-uniformity screens of a similar kind; the score adds the isolation criterion and a configurable, weighted combination of all criteria.

\subsection{Aligning and scaling each image from its own markers}\label{sec:calibration-results}

\begin{figure}[htb]
\centering
\panel{0.32\textwidth}{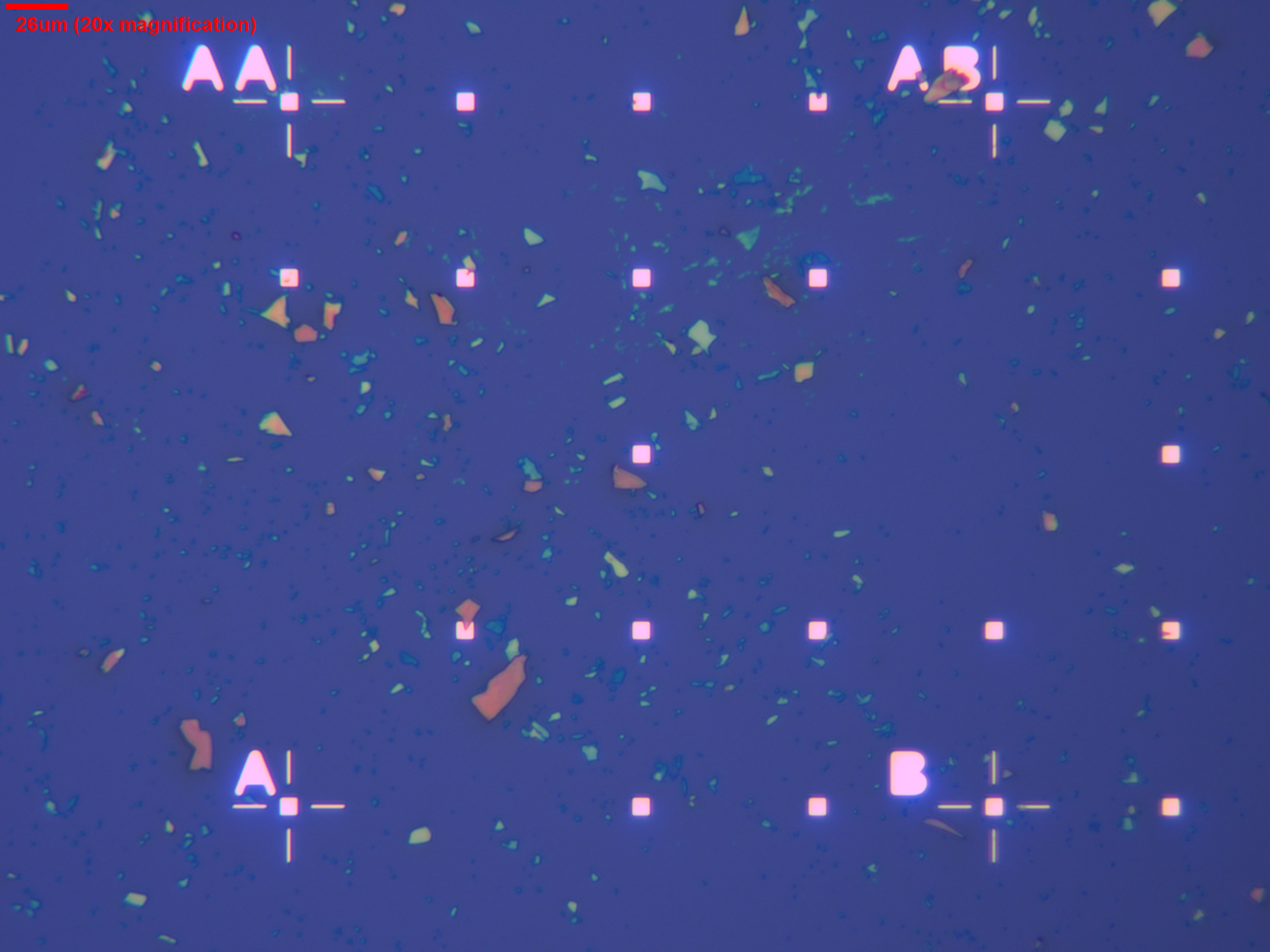}{\textbf{a.}}
\hfill
\panel{0.32\textwidth}{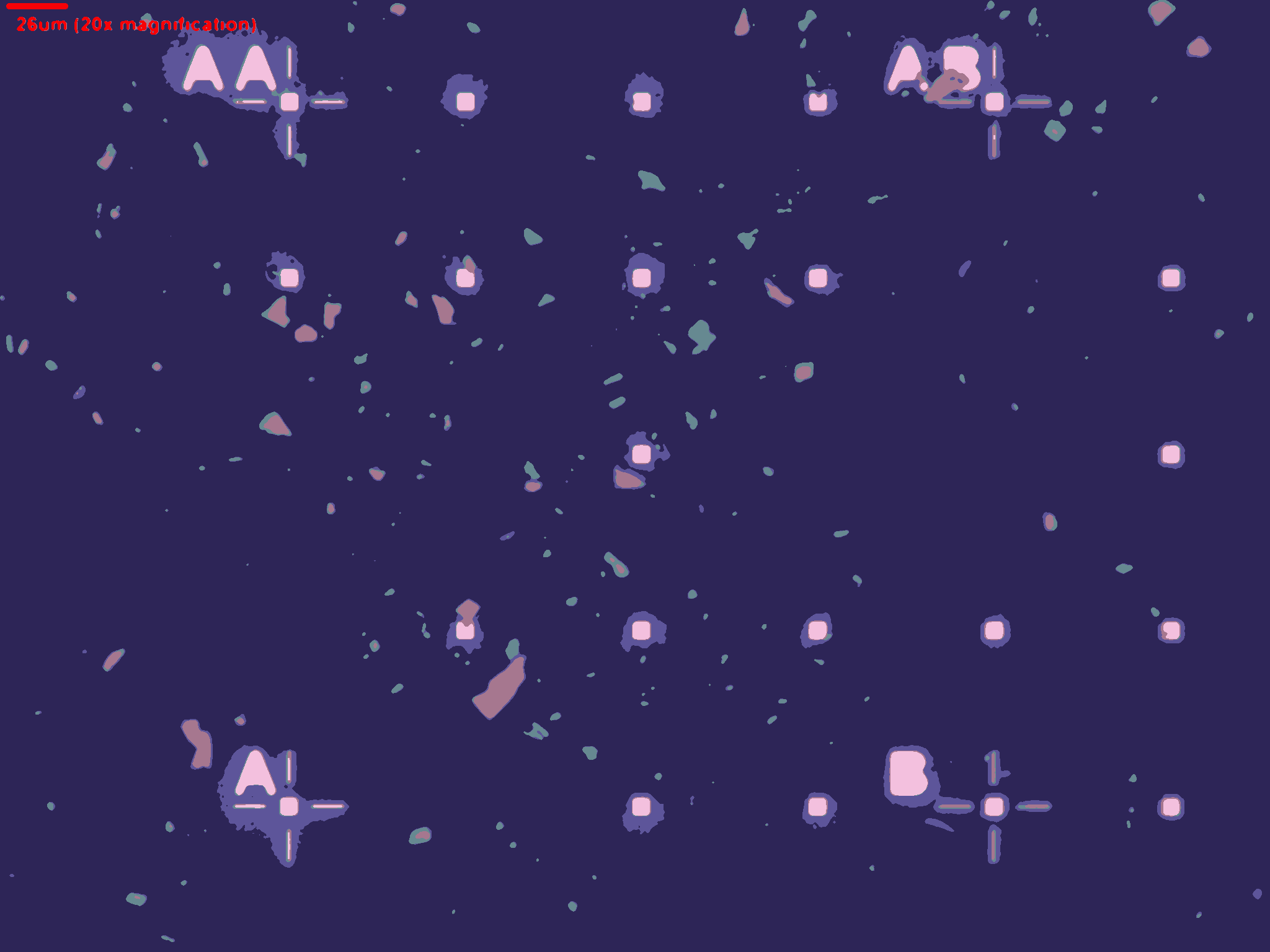}{\textbf{b.}}
\hfill
\panel{0.32\textwidth}{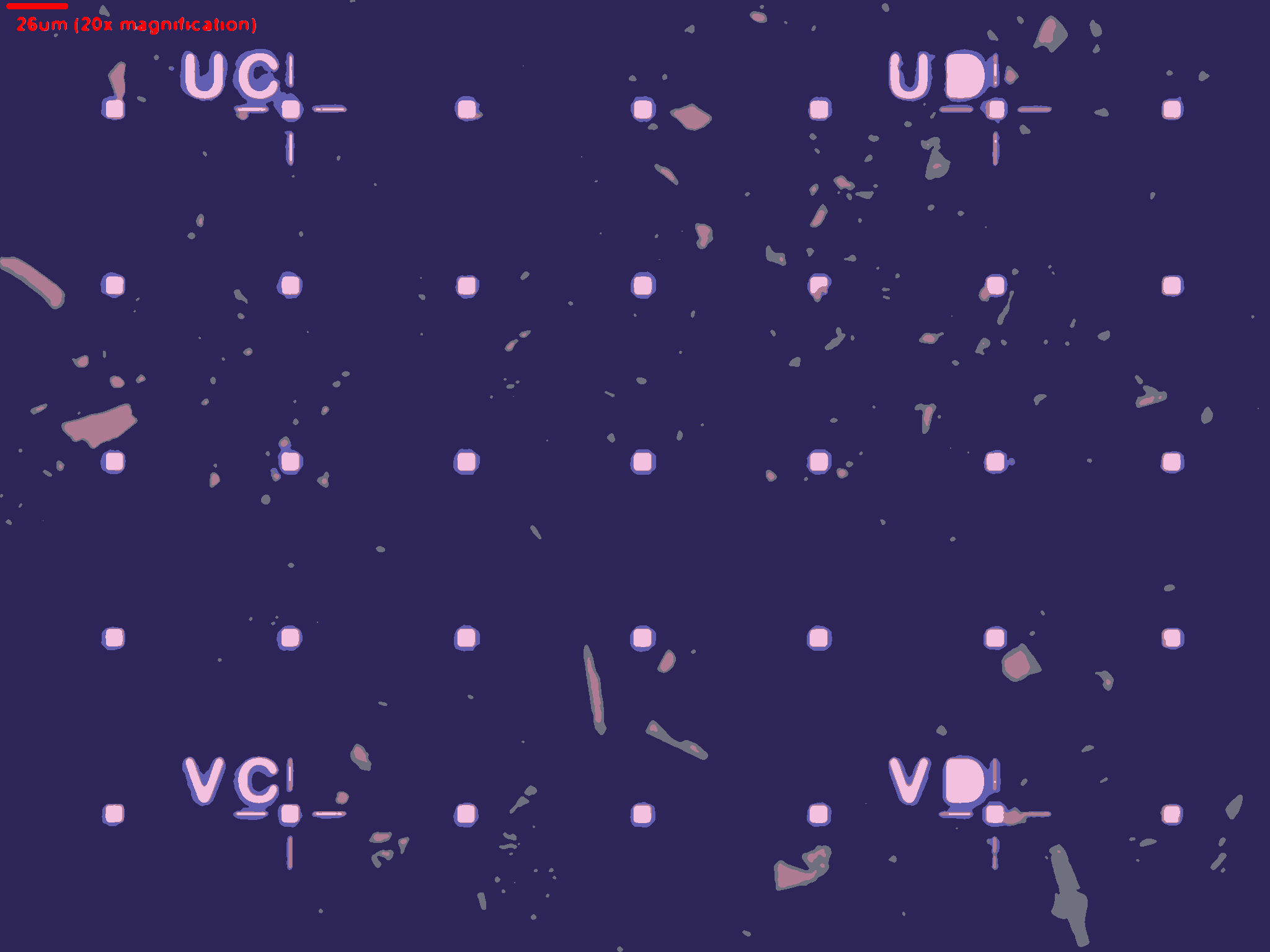}{\textbf{c.}}

\vspace{6pt}
\panel{0.32\textwidth}{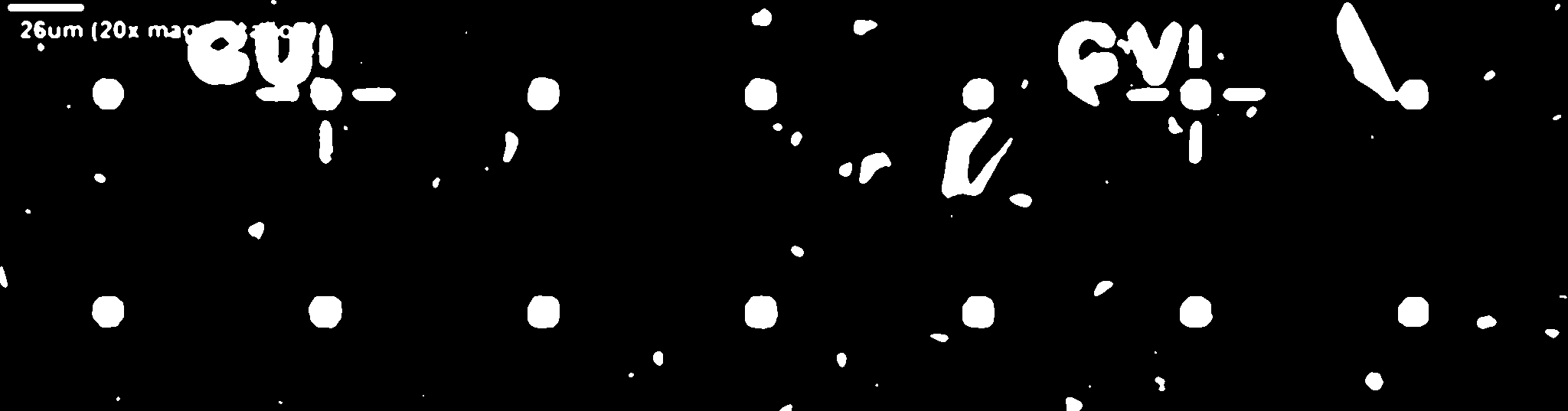}{\textbf{d.}}
\hspace{0.04\textwidth}
\panel{0.40\textwidth}{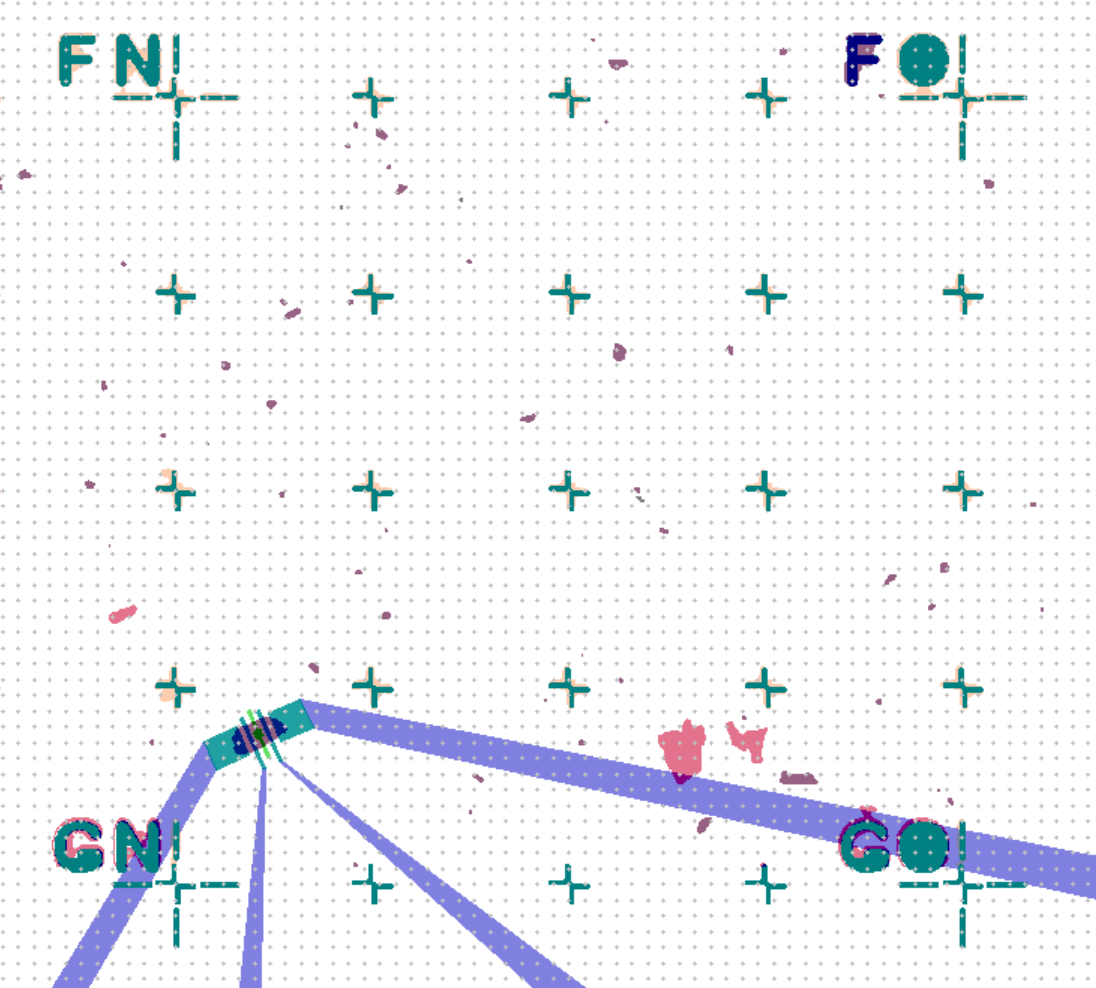}{\textbf{e.}}
\caption{\textbf{Alignment and scaling from the markers in each image.} \textbf{a} Raw optical image. \textbf{b} Its eight-colour reconstruction, keeping only the clusters the network judged to be chip content (flake material, crosshairs, grid dots). \textbf{c} Reconstruction of a second image. \textbf{d} Per-image Otsu brightness mask: both crosshairs, every grid dot and incidental flake material survive without any use of hue or saturation. \textbf{e} The pipeline on a chip with a different marker design, imaged on a different microscope, shown with a device already placed on the calibrated grid.}
\label{fig:alignment}
\end{figure}

Each image is first reduced to eight flat colours (Fig.~\ref{fig:alignment}a--c). Only some of these colour clusters are chip content (flake material, crosshairs and grid dots); the others are substrate or lighting variation. Which is which changes from image to image with illumination and flake thickness, so the kept clusters cannot be fixed in advance. A small convolutional neural network, trained on the selections an expert had already made when designing earlier chips by hand, chooses which clusters to keep. It judges each cluster by its spatial pattern rather than its colour and was trained with the cluster order randomly permuted, so the same network was used unchanged on every chip in this work (\hyperref[sec:m-reconstruction]{Methods}).

The two crosshair markers are then located in the reconstruction. Their appearance drifts with illumination from image to image, so no single colour threshold isolates them reliably. Two independent masks are therefore built: a substrate-exclusion mask in hue-saturation-value (HSV) space, and a per-image Otsu threshold\cite{otsu1979} of brightness that uses no colour information (Fig.~\ref{fig:alignment}d). Each mask is searched with two complementary strategies. Normalized cross-correlation\cite{lewis1995} with a template of the whole marker works when the marker is intact; a search for the marker's solid core, confirmed by evidence of its arms, tolerates arms broken up by colour drift. An optional third strategy recovers the centre from the arms alone, for markers whose core and one arm are missing from the image. Candidates from all strategies are scored on a common scale and duplicates are merged. The masks are tried one at a time and pooled only if neither succeeds alone, so spurious candidates from one mask cannot displace a clean result from the other. A marker pair is accepted only if it is unambiguous and its separation and orientation are consistent with the grid; otherwise the image is reported and left out rather than aligned on a guess, and it can be recovered in the review window. Size tolerances are expressed relative to the median area of the grid dots in each image, so the same configuration works across magnifications and cameras.

The two markers sit at known chip positions, so comparing their separation and direction in pixels with those on the chip gives the image's scale and rotation, and with them the on-chip position of every pixel (\hyperref[sec:m-alignment]{Methods}). Flake polygons are traced from the reconstruction by edge detection\cite{canny1986} and written, at their calibrated positions, into a working copy of the chip's master layout. Across the four chips, 272 of the 295 images in which detection kept at least one flake were aligned and placed in the layout; in the remaining 23 the markers could not be located.

The same scheme ran on a chip with a different marker design, imaged on another microscope whose settings change the apparent colour of the flakes, the substrate and the markers (Fig.~\ref{fig:alignment}e). Adapting it required a new marker template and a few configuration values, and the arm-triangulation strategy was switched off because these markers have no comparable arms. The downstream stages, which see only calibrated coordinates, were unaffected. A different naming scheme would require only a new decoder for the image name, provided the name still encodes the image's position on the chip.

\subsection{Device layouts fitted to each flake and routed across the chip}\label{sec:placement-results}

A flake contour has no canonical orientation and no straight edge to reference, so the device layout stage reduces it to one reference rectangle in the flake's own orientation and defines every device element as an offset from that rectangle. The orientation is that of the contour's minimum-area bounding rectangle\cite{toussaint1983}. For a Hall bar, the reference is the largest rectangle inscribed in the contour, since the channel must lie entirely on the flake. For an LP ring, the reference is the bounding rectangle itself, because the ring is etched out of the flake rather than fitted inside it (\hyperref[sec:m-geometry]{Methods}).

\begin{figure}[htb]
\centering
\panel{0.48\textwidth}{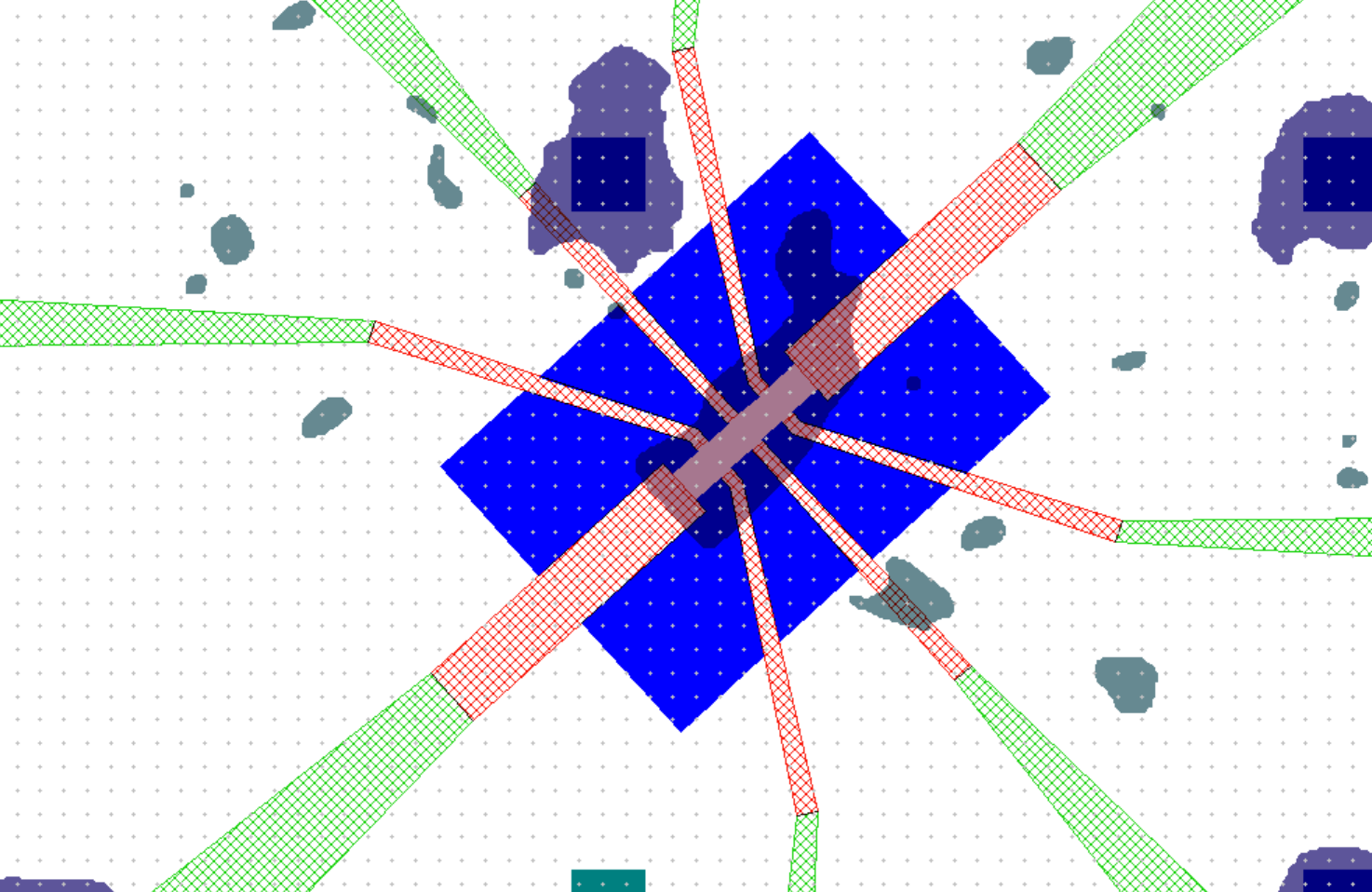}{\textbf{a.}}
\hfill
\panel{0.48\textwidth}{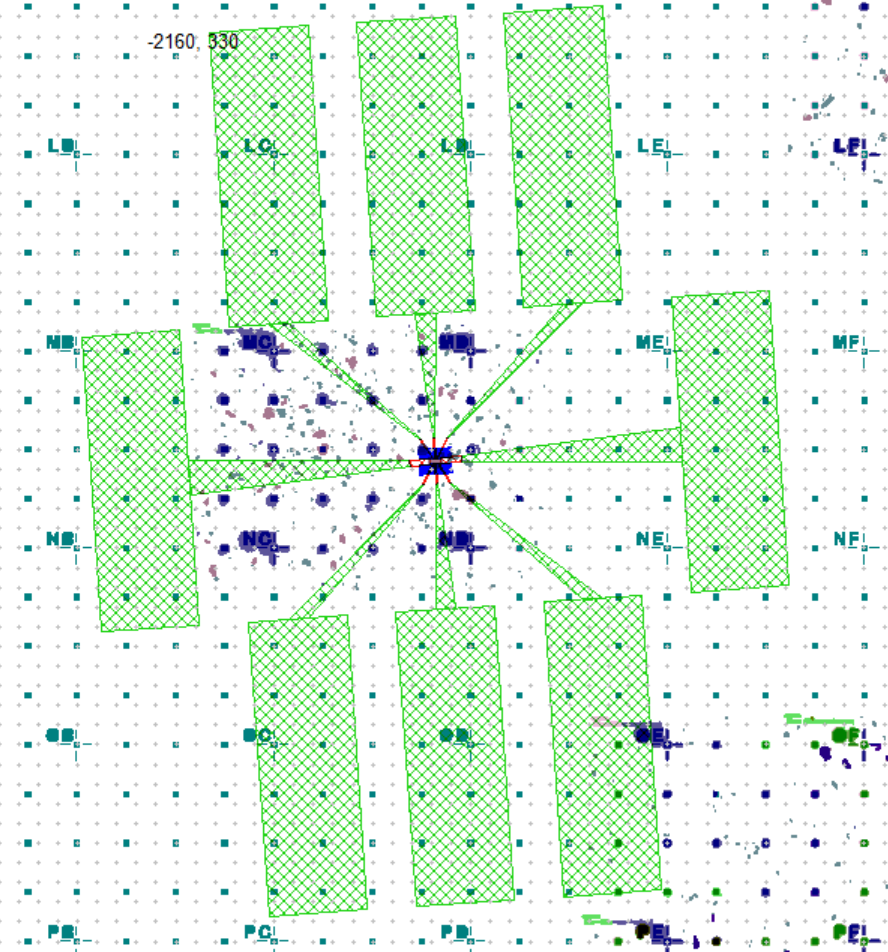}{\textbf{b.}}

\vspace{6pt}
\panel{0.32\textwidth}{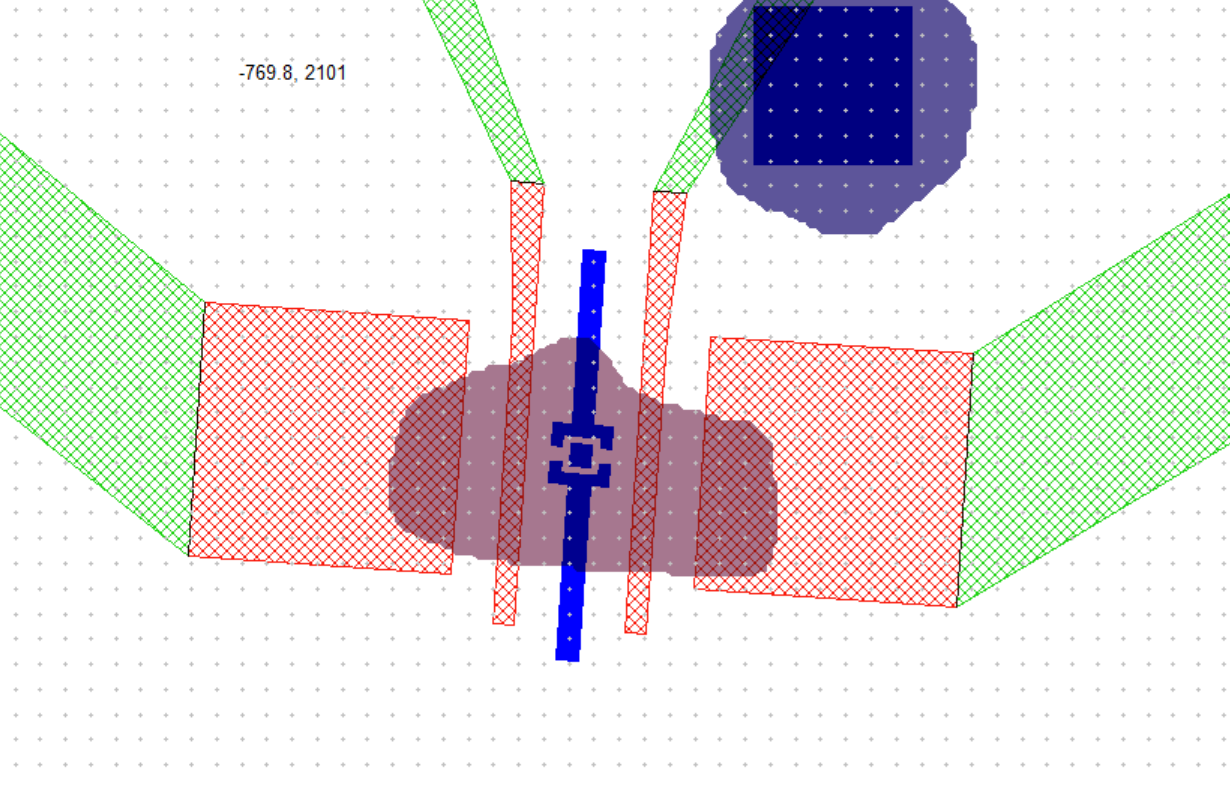}{\textbf{c.}}
\hfill
\panel{0.32\textwidth}{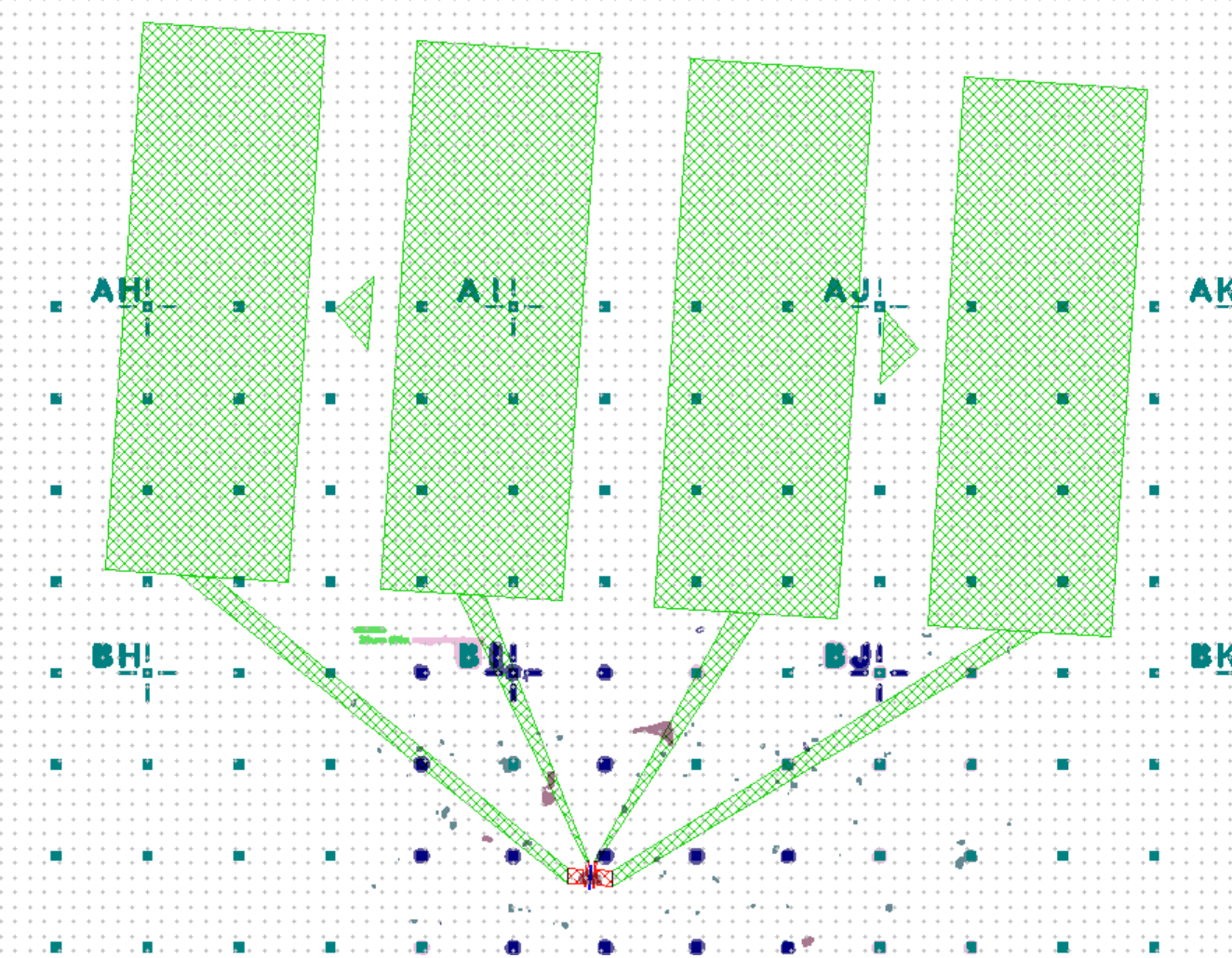}{\textbf{d.}}
\hfill
\panel{0.32\textwidth}{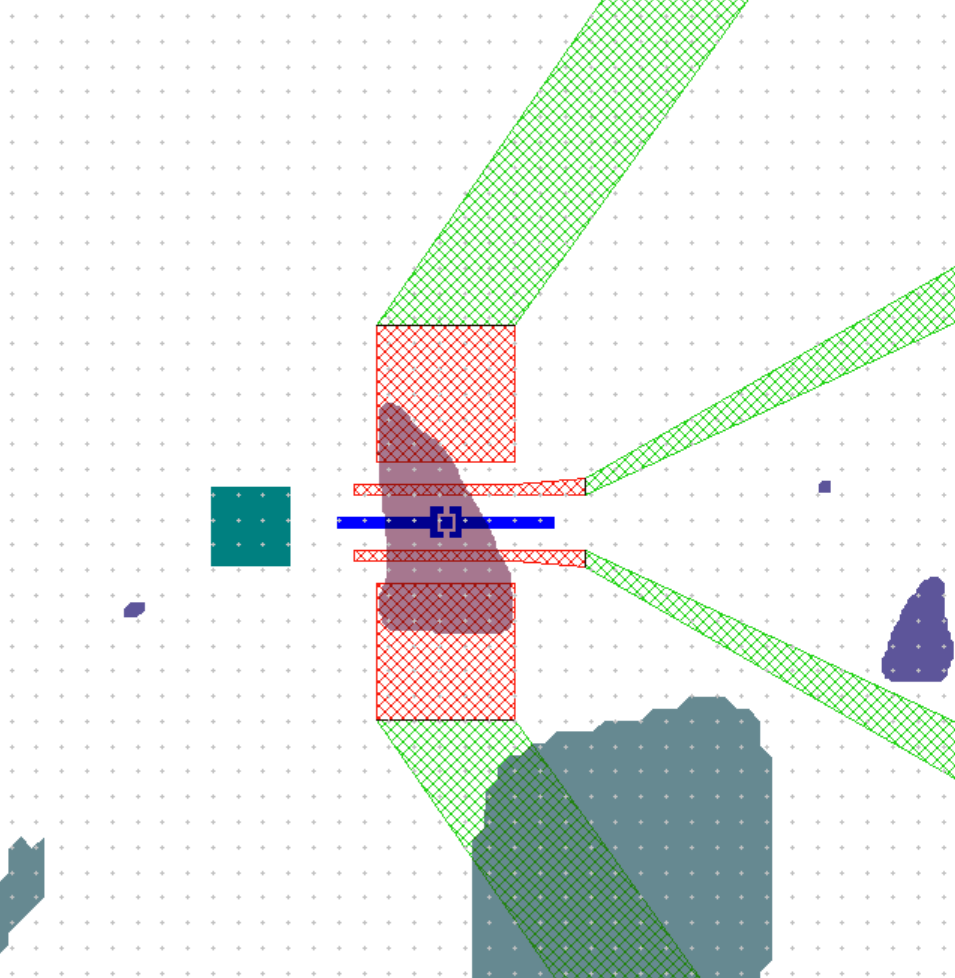}{\textbf{e.}}
\caption{\textbf{Device geometry fitted to the flake contour.} Exported CIF geometry viewed in mask layout software (CleWin). \textbf{a} Hall-bar contacts, two current and six voltage, fitted to the flake's largest inscribed rectangle. \textbf{b} The eight bond pads they are routed to. \textbf{c} LP ring contacts, two current and two voltage, around the etched ring. \textbf{d} The four bond pads of the ring device. \textbf{e} A second LP ring device, with voltage contacts widening toward their traces. Blue, etch mask; red, contacts (fine EBL exposure); green, traces and bond pads (coarse EBL exposure).}
\label{fig:geometry}
\end{figure}

Every other dimension follows from the reference rectangle through configuration fields (\hyperref[sec:m-geometry]{Methods}). A Hall bar is defined by its etch mask, which leaves a bar of flake: a rectangular channel within the reference rectangle, with six voltage probes extending from its long sides, three per side at the quarter, half and three-quarter positions. Two current contacts span the ends of the channel, and a voltage contact covers each probe. Each voltage contact widens toward its outer end, and the four outer ones bend away from the centre pair. An LP ring is also defined by its etch mask, which leaves a square loop of flake around a square hole, joined to the rest of the flake by a lead on each of two opposite sides; it has two current and two voltage contacts placed symmetrically along its length. The contacts and the bond pads with their traces are written on separate metallization layers, because they need different lithography settings, and the etch masks on a third layer (Fig.~\ref{fig:geometry}).

\begin{figure}[htb]
\centering
\panel{0.32\textwidth}{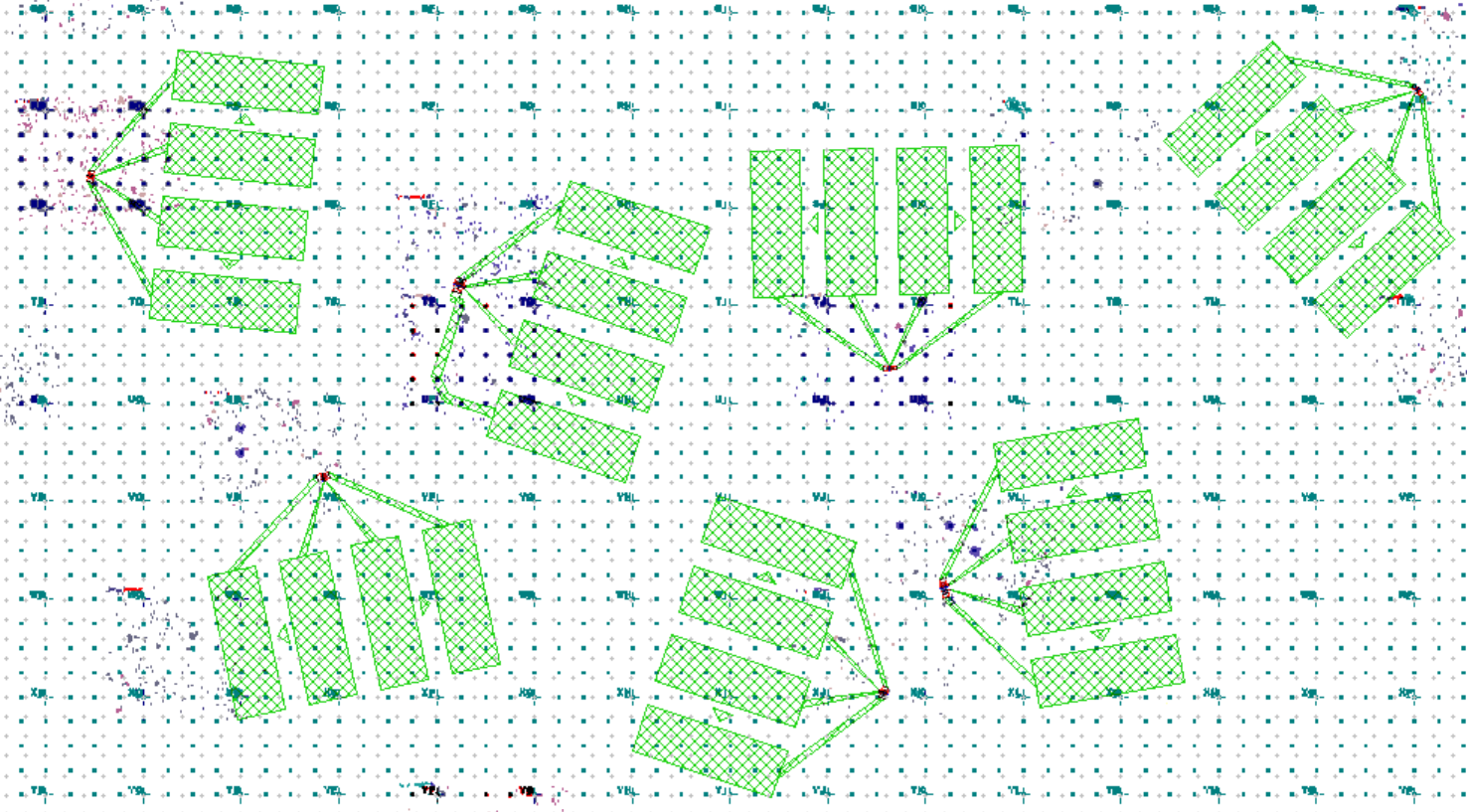}{\textbf{a.}}
\hfill
\panel{0.32\textwidth}{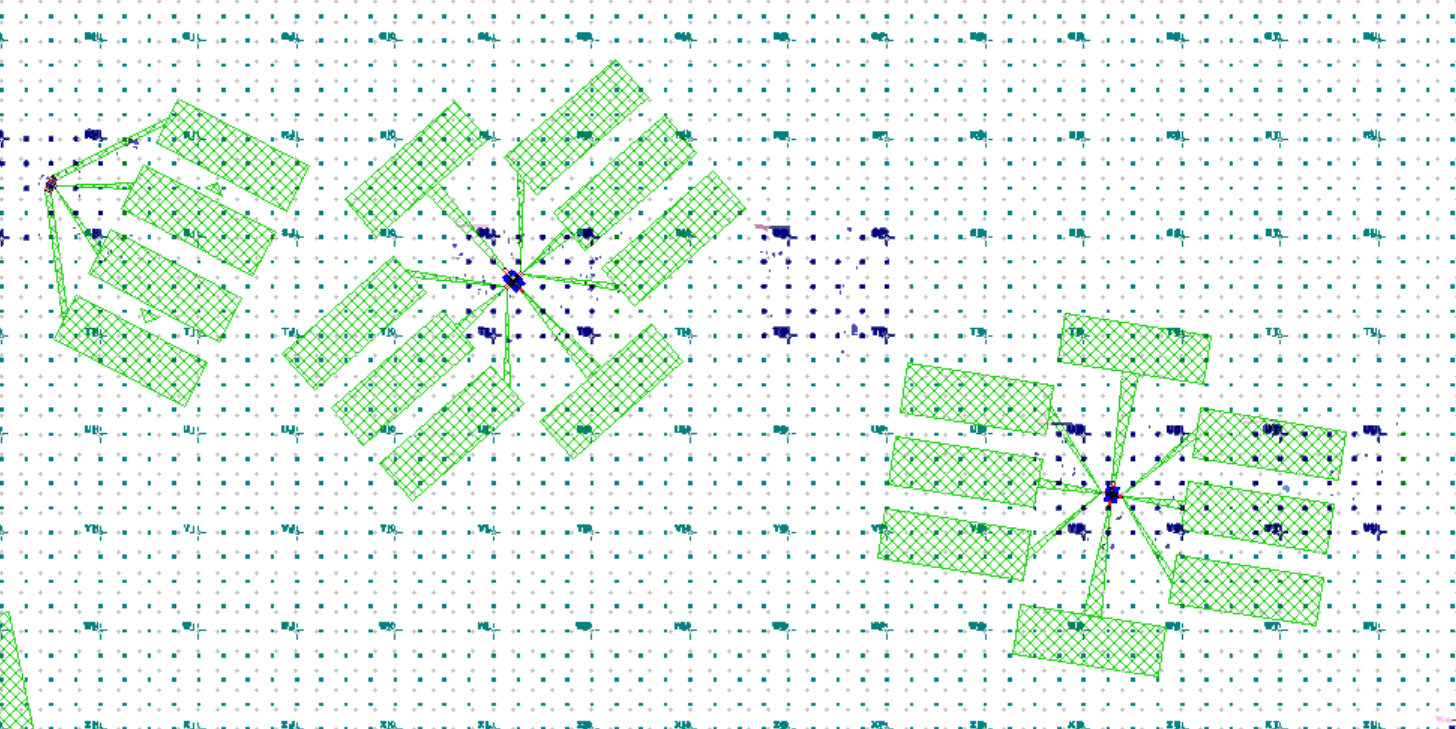}{\textbf{b.}}
\hfill
\panel{0.32\textwidth}{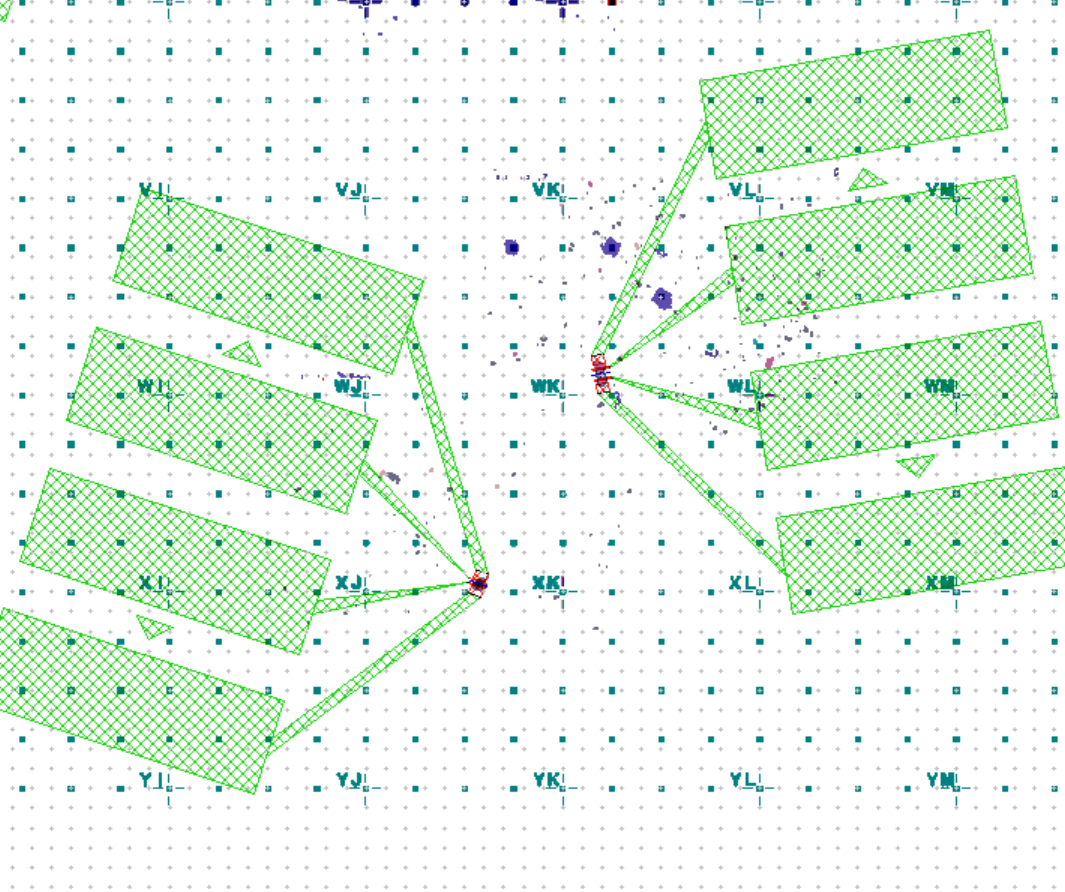}{\textbf{c.}}
\caption{\textbf{Collision-free placement across the chip.} Exported CIF geometry for devices placed independently on different flakes. \textbf{a} Several LP ring devices in one chip region. \textbf{b} An LP ring (left, four pads) beside two Hall bars (eight pads each). \textbf{c} Two devices close enough that their arms interleave without touching. Each device is checked against all geometry written before it.}
\label{fig:chipscale}
\end{figure}

Routing is shared by both device types. Each device reports only where its contacts end, in which direction and how wide; a common routine fits the bond pads into a square work area around the device and connects every contact to its bond pad with a straight, tapered trace. All placed geometry accumulates in one occupied region for the whole chip, which also contains the other selected flakes, so a later device is routed around every earlier one. A device that would collide with existing geometry or cross the chip-border margin is abandoned entirely rather than written in part (Fig.~\ref{fig:chipscale}). The pipeline places up to the number of devices the user requests; the number actually placed is limited by the available flakes and the chip area their devices occupy.

\subsection{Optional review checkpoints}\label{sec:review-results}

\begin{figure}[htb]
\centering
\panel{0.48\textwidth}{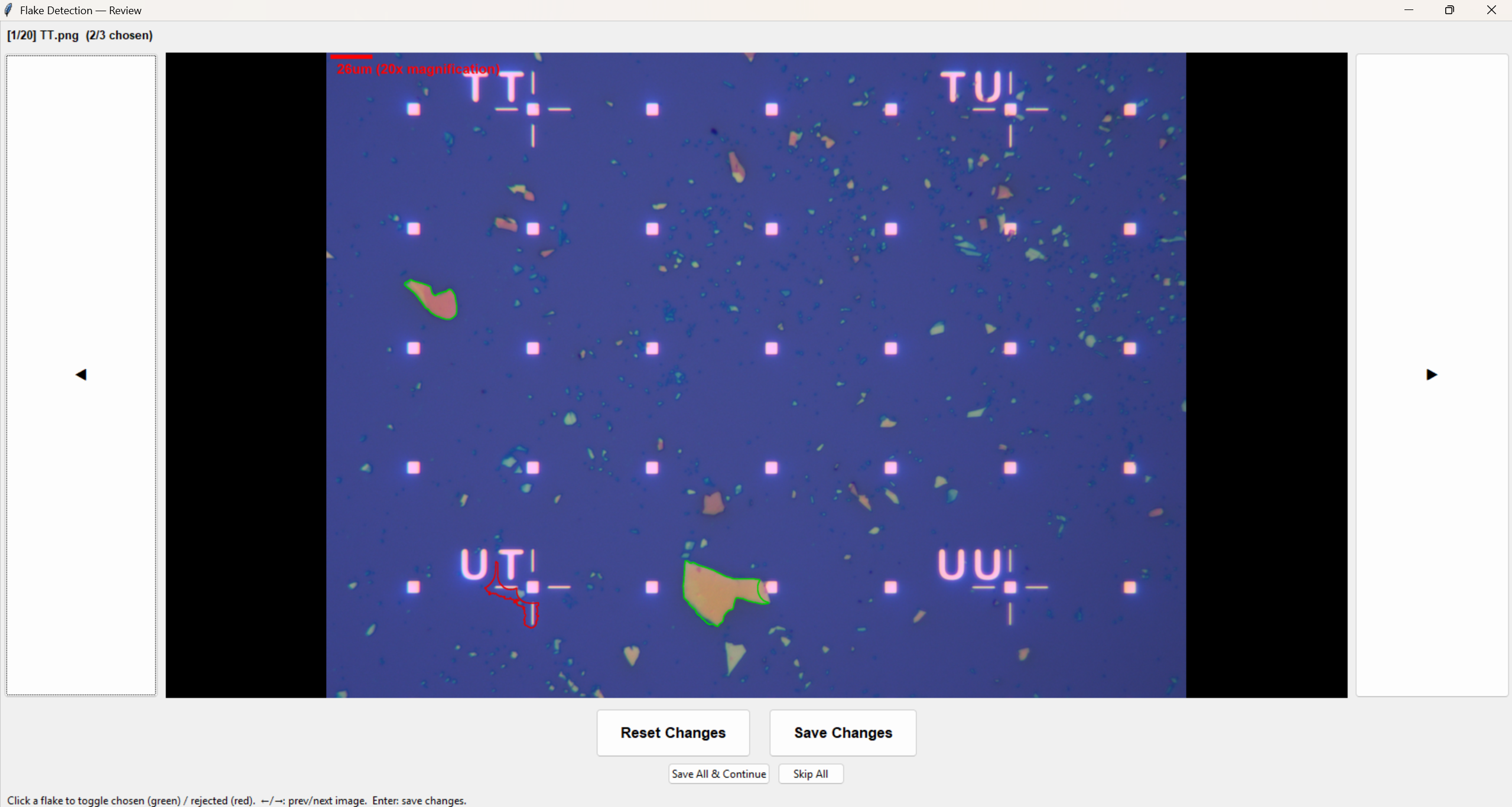}{\textbf{a.}}
\hfill
\panel{0.48\textwidth}{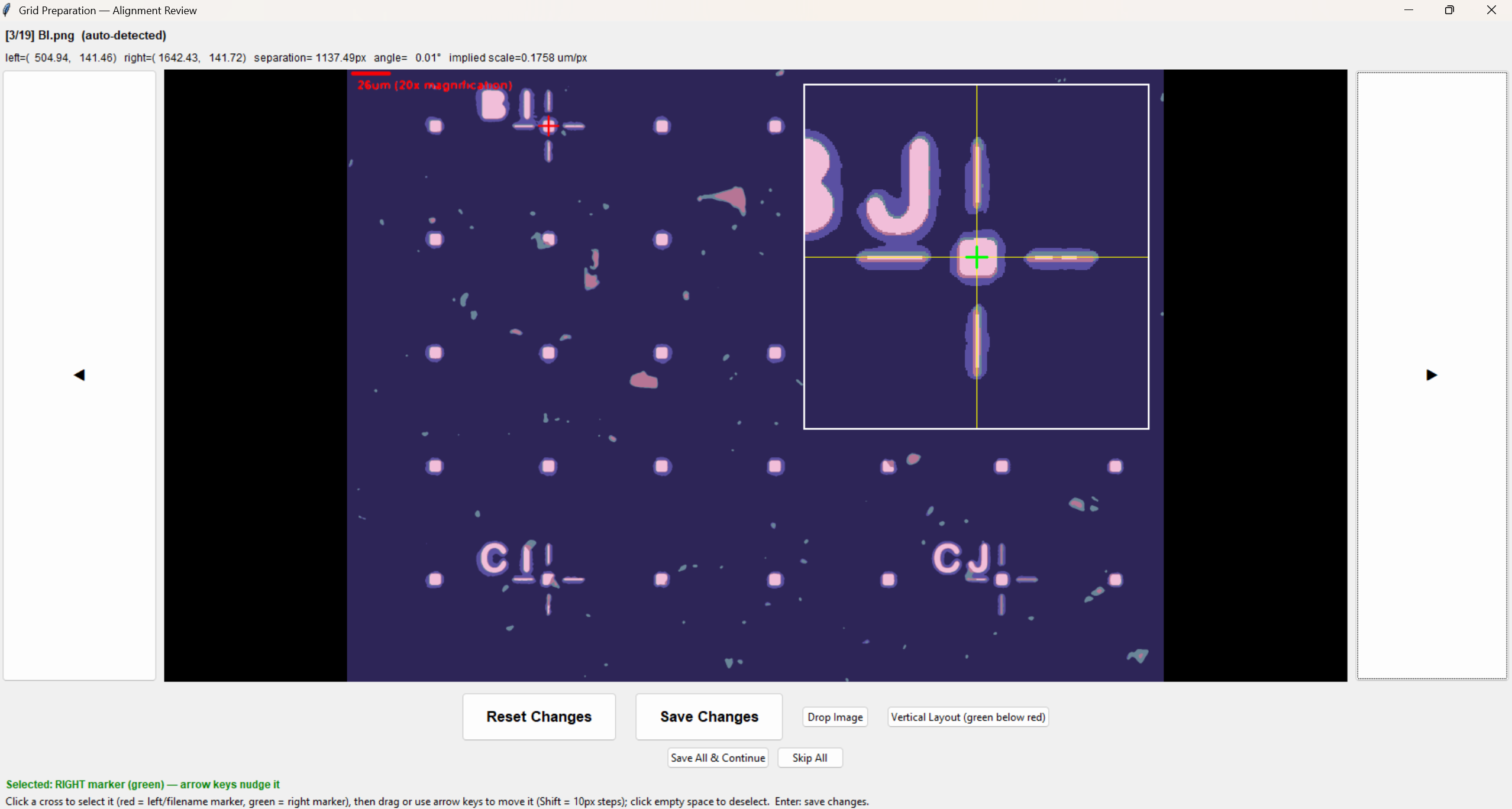}{\textbf{b.}}

\vspace{6pt}
\panel{0.24\textwidth}{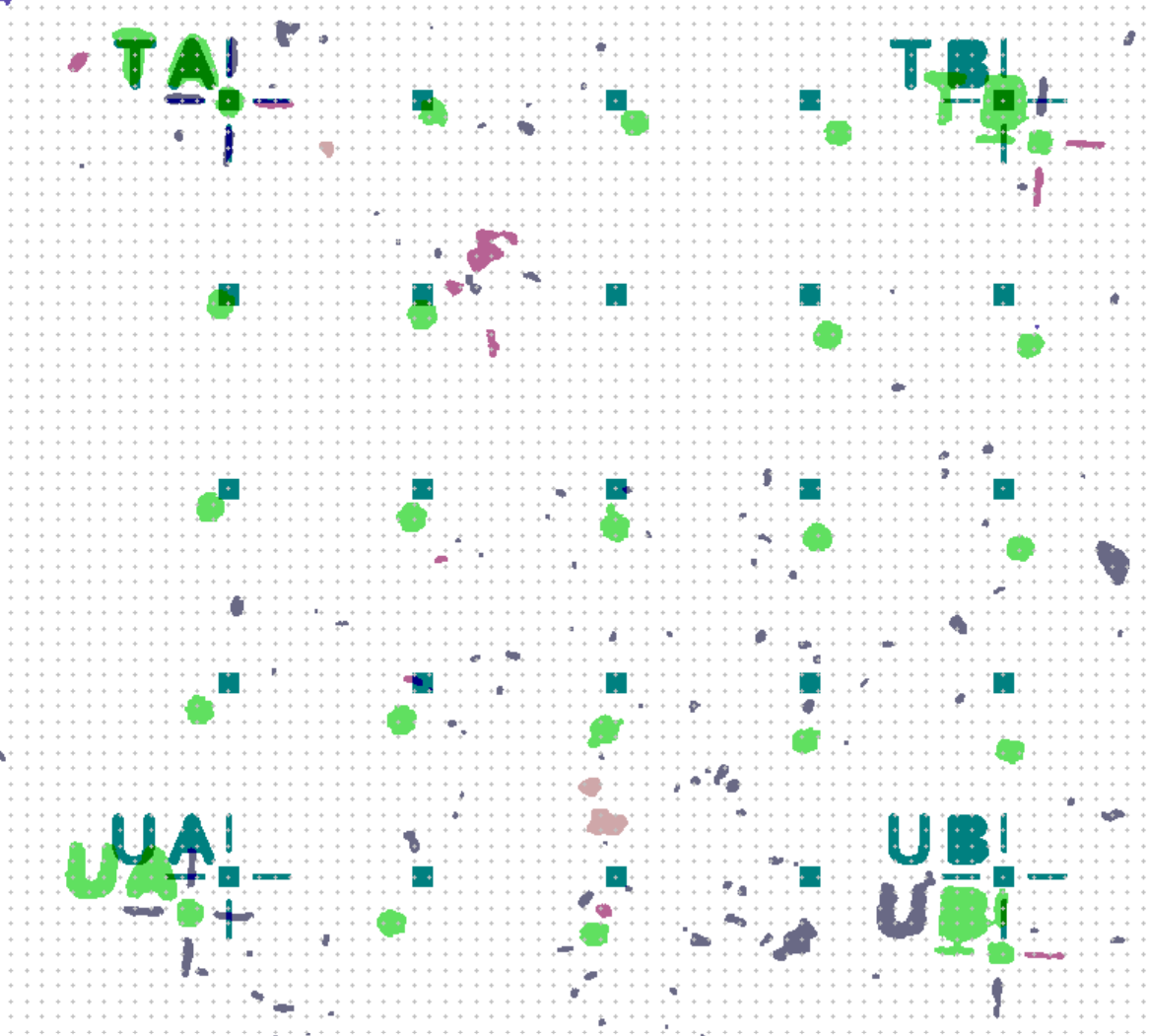}{\textbf{c.}}
\hfill
\panel{0.24\textwidth}{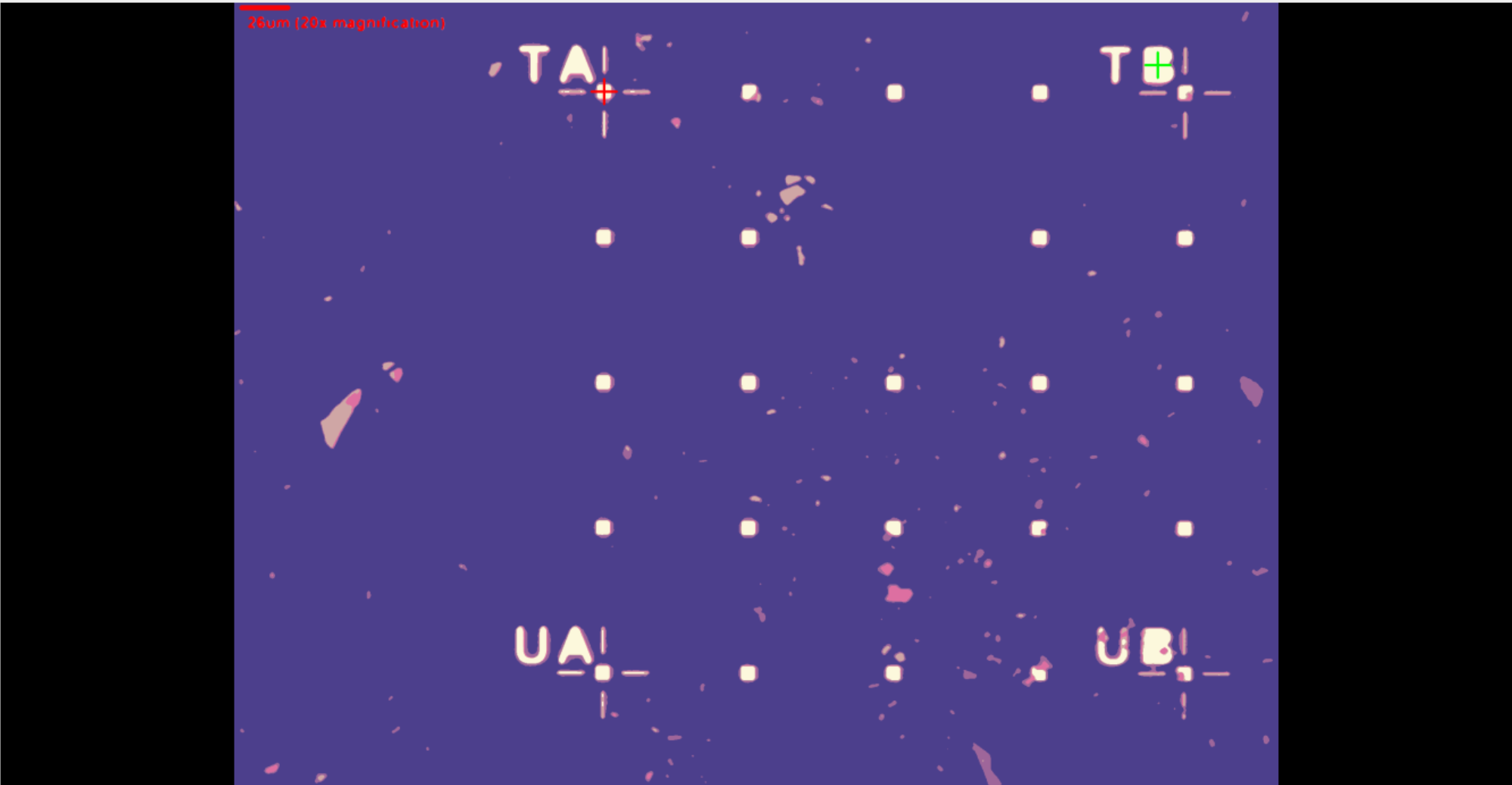}{\textbf{d.}}
\hfill
\panel{0.24\textwidth}{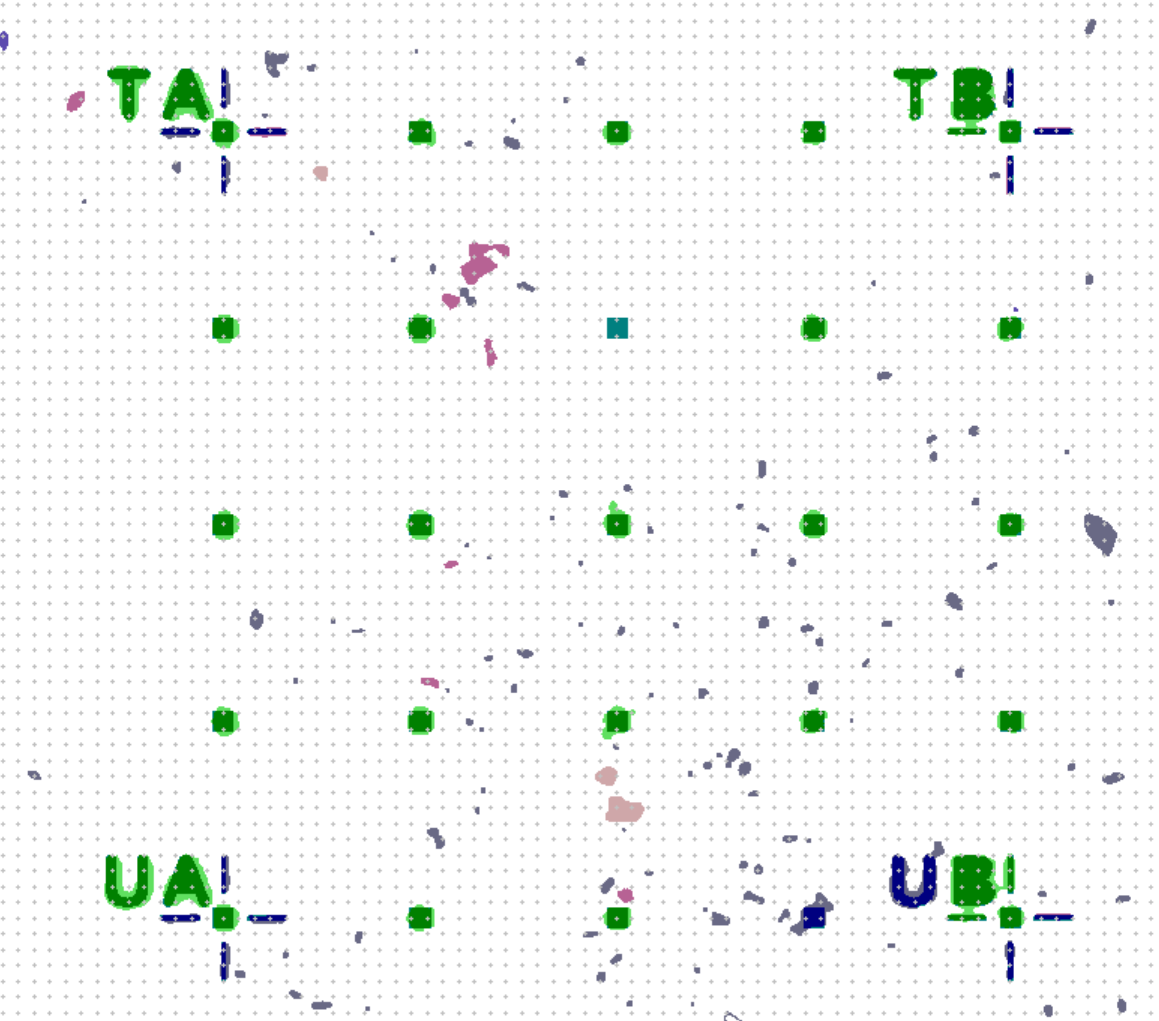}{\textbf{e.}}
\hfill
\panel{0.24\textwidth}{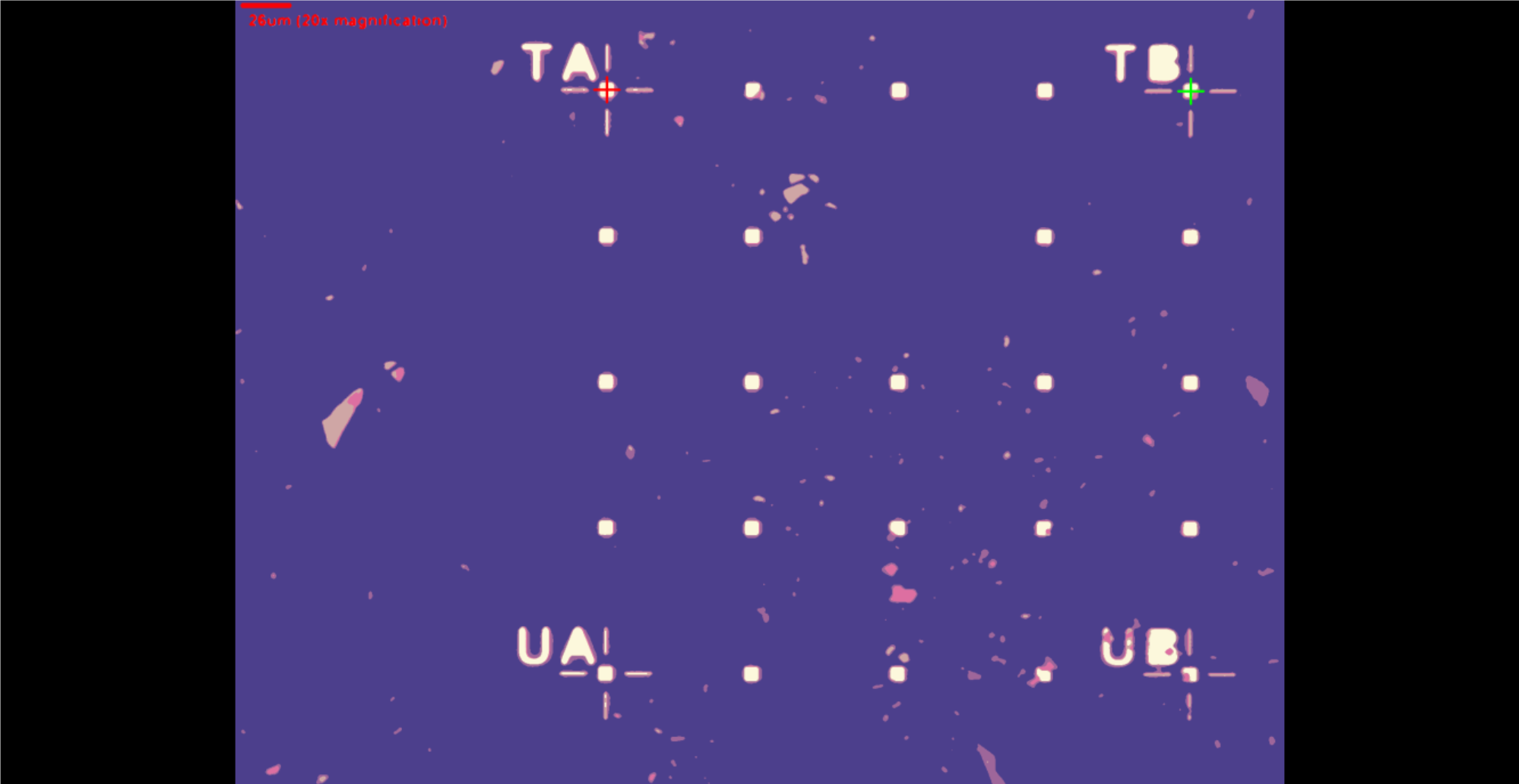}{\textbf{f}}
\caption{\textbf{Review checkpoints.} \textbf{a} Detection review: every candidate that passed the size filter is outlined as kept or rejected and can be toggled with a click. \textbf{b} Alignment review, with draggable crosshairs, magnifier and live scale readout. \textbf{c},~\textbf{d} A real detection error, the right crosshair placed off the marker centre, in the detection output (\textbf{c}) and in the review window (\textbf{d}). \textbf{e},~\textbf{f} A correctly detected image for comparison. }
\label{fig:review}
\end{figure}

Flake detection, alignment and scaling can each pause for inspection by the user before their output is used (Fig.~\ref{fig:review}). The detection review outlines every candidate that passed the size filter, kept or rejected, and a click toggles its state, so a false positive can be removed and a narrowly missed flake recovered. The alignment review shows both detected crosshairs, which can be dragged to corrected positions with the help of a magnifier and a live readout of marker separation, angle and implied scale. An image can also be dropped or, when no horizontally adjacent marker pair is in view, assigned a vertically adjacent pair, so that a flake outside the usual image framing can still be used. In normal use, the review caught a crosshair placed visibly off its marker (Fig.~\ref{fig:review}c,d), an error that would have distorted every polygon and contact on that image and that one drag corrected before the transform was computed. The device layout stage has no review window; its output is inspected and edited in mask layout software.

\subsection{Full-pipeline runs and comparison with domain expert}\label{sec:pipeline-runs}

We ran the whole pipeline once, unattended, on each of four chips of exfoliated 4Hb-TaS$_2$: Chip A, Chip B, Chip C  and Chip D (Table~\ref{tab:pipeline-runs}). Chip C was configured for LP rings only. Independently, a domain expert laid out each chip by hand from the same images, aiming to fit as many devices as possible. Across the 397 images, the pipeline placed 119 devices and the expert 168.

\begin{table}[htb]
\centering
\small
\begin{tabular}{@{}llrrrrrr@{}}
\toprule
Chip (images) & & Img/flake & Img/CIF & Flakes & Devices & Hall & Ring \\
\midrule
\multirow{2}{*}{A (77)} & Pipeline & 68 & 64 & 159 & 27 & 5 & 22 \\
 & DE & -- & -- & -- & 46 & 4 & 42 \\
\midrule
\multirow{2}{*}{B (134)} & Pipeline & 104 & 95 & 184 & 32 & 3 & 29 \\
 & DE & -- & -- & -- & 47 & 3 & 44 \\
\midrule
\multirow{2}{*}{C (139)} & Pipeline & 91 & 82 & 137 & 41 & 0 & 41 \\
 & DE & -- & -- & -- & 47 & 0 & 47 \\
\midrule
\multirow{2}{*}{D (47)} & Pipeline & 32 & 31 & 43 & 19 & 2 & 17 \\
 & DE & -- & -- & -- & 28 & 7 & 21 \\
\bottomrule
\end{tabular}
\caption{\textbf{Full-pipeline runs on four chips.} Img/flake, images with at least one selected flake; Img/CIF, images aligned and placed in the chip layout; Flakes, flakes selected by detection; Devices, selected flakes on which a device was placed, split into Hall bars and LP rings. DE, the domain expert's manual layout of the same chip.}
\label{tab:pipeline-runs}
\end{table}

\begin{figure}[htb]
\centering
\includegraphics[width=\textwidth]{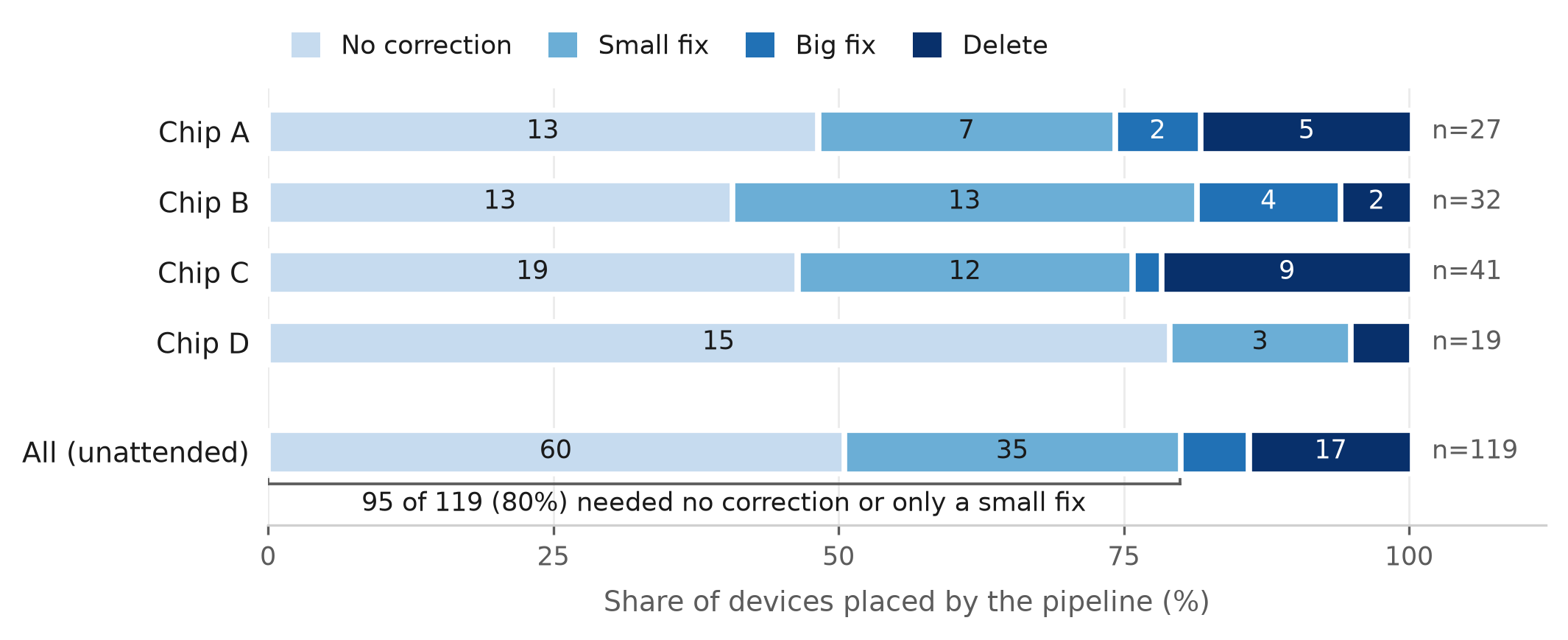}
\caption{\textbf{Expert review of the devices placed by the pipeline.} Share of each run's devices that the expert left unchanged, corrected with a small fix (up to about a minute), corrected with a big fix (reshaping some or all contacts), or deleted; counts appear in each segment and run totals at the right. Where a device received more than one mark, deletion takes precedence over a big fix and a big fix over a small one. The aggregate row covers all four runs.}
\label{fig:fixburden}
\end{figure}

The expert then inspected every device the pipeline placed and recorded whether it was usable as exported, needed a small fix (up to about one minute of manual editing), needed some or all of its contacts reshaped (a big fix) or should be deleted (\hyperref[sec:m-expert]{Methods}). Of the 119 devices, 60 needed no correction, 35 a small fix, 7 a big fix and 17 were deleted (Fig.~\ref{fig:fixburden}). A deletion does not always mean an unsuitable flake: it also covers devices not worth fixing, crowded sites and flakes passed over for a better neighbour. Six devices carried alignment errors from a mislocated marker; five of these fell in the small-fix group and one among the deletions.

Comparing the two layouts shows how a DE and the pipeline choose from the same pool of candidates (Table~\ref{tab:de-agreement}). The pipeline placed devices on 59 of the 168 flakes the expert used, and on 60 further flakes the expert did not use. Together, the two layouts cover 228 flakes, more than either alone: a chip usually offers more candidate flakes than any single layout uses, so two sound selections can differ. For 34 of the expert's flakes that the pipeline left unused, the expert recorded whether leaving them out was justified. In 24 cases it was: the expert was unsure of the flake, its colour lay outside the configured range, or the expert had built a device on only part of the flake, which the pipeline does not attempt. The remaining differences follow from how the pipeline chooses. It fits devices as rectangles with fixed size and spacing rules, whereas the expert shapes each device freely. It places devices one at a time in order of score, so it may take one high-scoring flake where the expert fits two lower-scoring ones. Its thresholds reflect one configuration rather than the expert's preferences, and can be adjusted. The devices on flakes that only the pipeline chose are included in the review above (Fig.~\ref{fig:fixburden}).

\begin{table}[htb]
\centering
\small
\begin{tabular}{@{}lrrrrrr@{}}
\toprule
Chip & DE picks & Also placed & DE only & Pipeline only & Justified & Unjustified \\
\midrule
A & 46 & 15 & 31 & 12 & 7 & 3 \\
B & 47 & 11 & 36 & 21 & 8 & 2 \\
C & 47 & 18 & 29 & 23 & 7 & 3 \\
D & 28 & 15 & 13 &  4 & 2 & 2 \\
\midrule
Total & 168 & 59 & 109 & 60 & 24 & 10 \\
\bottomrule
\end{tabular}
\caption{\textbf{Agreement on flake selection.} Of the flakes the domain expert selected, those on which the pipeline also placed a device and those it did not; flakes only the pipeline used; and the expert's verdicts on the pipeline's rejections where one was recorded.}
\label{tab:de-agreement}
\end{table}

\subsection{Effect of reviewing during the run}\label{sec:reviewed-run}

The review checkpoints address two kinds of correction before any geometry is written, detection and alignment errors, and give the user control over which flakes become devices. At the detection checkpoint, the user can remove poorly chosen flakes and keep flakes the detector rejected. Most deletions can be avoided in this way, by judging each flake and its surroundings before any device is drawn. The user can also favour particular flakes without changing any score: because devices are placed one at a time and each occupies chip area and routing space, deselecting a good flake that would claim space near a preferred one leaves that space for the preferred flake. At the alignment checkpoint, an alignment error is fixed by dragging the two markers, after which the transform and every contact derived from it are recomputed. Together, deletions and alignment errors account for 22 of the 59 corrected devices: the 17 deletions and the five alignment errors among the small fixes. Most of the remaining corrections are small geometric adjustments to the exported layout, and many stem from conditions that are hard to express as rules: a trace that crosses material the pipeline did not select, such as unselected flakes or the chip's own patterned features; a crack whose effect depends on where it lies; or a site too crowded to be worth using. A researcher recognizes these at a glance, but they are corrected in mask layout software rather than at a checkpoint.

To measure what human review recovers, we ran Chip A and Chip C again with both checkpoints enabled. A user corrected each stage before its output passed on (Table~\ref{tab:reviewed}). On both chips the reviewed run left fewer devices to correct and almost none to delete. On Chip A, review raised the number of aligned images from 64 to 70 and the number of placed devices from 27 to 34; 24 of these needed no correction, against 13 unattended, and deletions fell from 5 to 1. On Chip C both runs placed 41 devices, though not all on the same flakes; 32 needed no correction against 19, deletions fell from 9 to none, and corrections of every kind fell from 22 to 9. Agreement with the expert's own selection rose on both chips, from 15 to 19 of 46 picks on Chip A and from 18 to 28 of 47 on Chip C. The reviewed runs took two to three times as long as the unattended ones, 26.4 against 12.4 minutes on Chip A and 21.7 against 8.4 minutes on Chip C, including the time the user spent at both checkpoints.

\begin{table}[htb]
\centering
\small
\begin{tabular}{@{}lrrrr@{}}
\toprule
 & \multicolumn{2}{c}{Chip A} & \multicolumn{2}{c}{Chip C} \\
\cmidrule(lr){2-3}\cmidrule(lr){4-5}
 & Unattended & Reviewed & Unattended & Reviewed \\
\midrule
Images aligned & 64 & 70 & 82 & 76 \\
Devices placed & 27 & 34 & 41 & 41 \\
No correction & 13 & 24 & 19 & 32 \\
Small fix & 7 & 9 & 12 & 8 \\
Big fix & 2 & 0 & 1 & 1 \\
Deleted & 5 & 1 & 9 & 0 \\
\midrule
Usable with at most a small fix & 20 & 33 & 31 & 40 \\
Corrections required & 14 & 10 & 22 & 9 \\
Agreement with expert & 15/46 & 19/46 & 18/47 & 28/47 \\
Runtime (min) & 12.4 & 26.4 & 8.4 & 21.7 \\
\bottomrule
\end{tabular}
\caption{\textbf{Two chips run unattended and with review.} Categories as in Fig.~\ref{fig:fixburden}. Images aligned counts images placed in the chip layout. Corrections required counts every device that needed a small fix, a big fix or deletion. Agreement counts the expert's selected flakes on which the pipeline also placed a device.}
\label{tab:reviewed}
\end{table}

\subsection{Runtime}\label{sec:runtime}

\begin{figure}[htb]
\centering
\includegraphics[width=\textwidth]{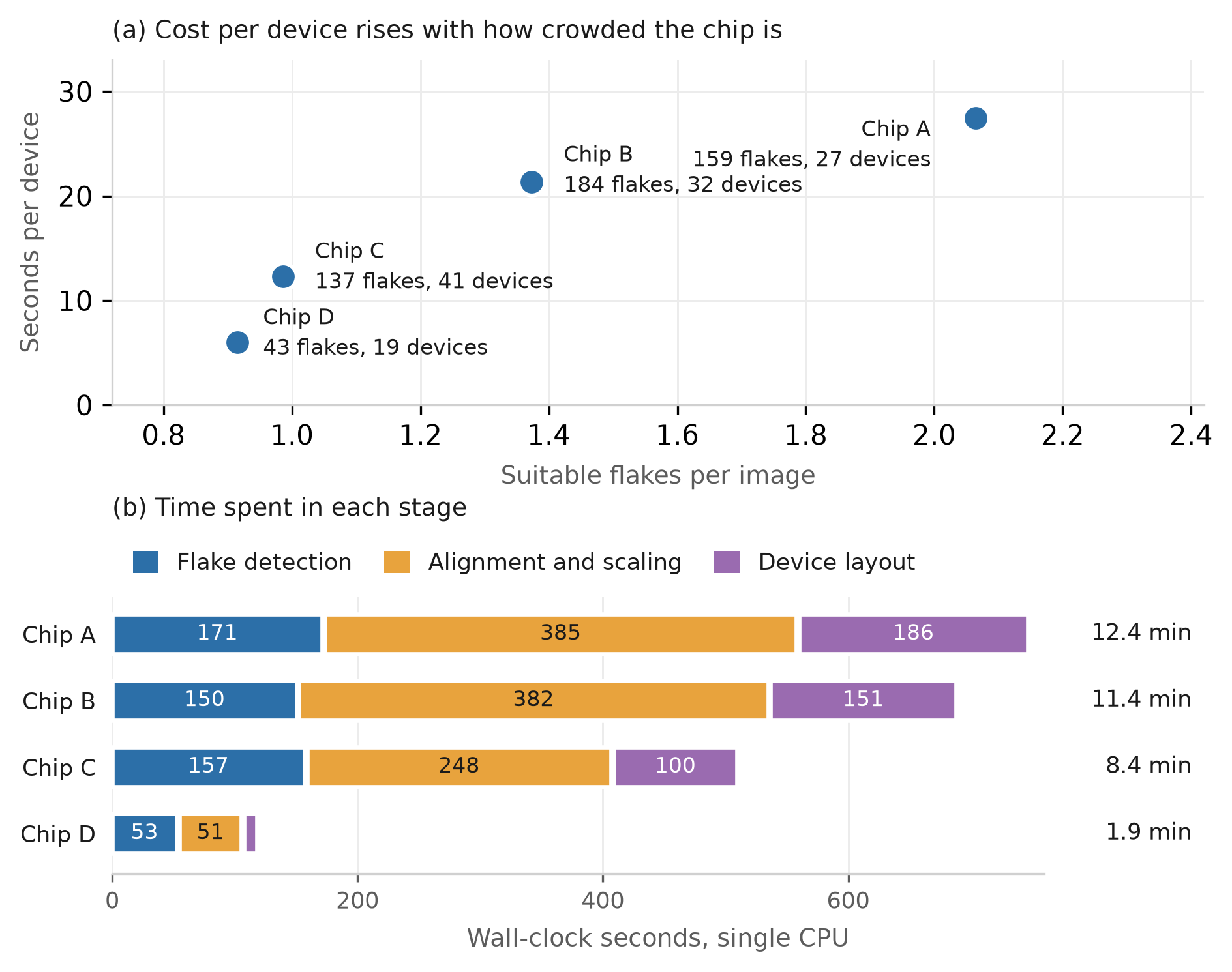}
\caption{\textbf{Runtime of the unattended runs.} Measured on a laptop (Intel Core i7-12700H, 32~GB memory, NVIDIA RTX 4060 Laptop GPU) from each run's own timing record. \textbf{a} Cost per device against the number of selected flakes per image;  \textbf{b} Total runtime split by stage; stage times fall 1--3~s short of the run total, the remainder being orchestration overhead.}
\label{fig:runtime}
\end{figure}

The four unattended runs together read 397 images and wrote 119 device layouts in 34 minutes on a consumer laptop, with no user present (Fig.~\ref{fig:runtime}). Flake detection and device layout run on the central processing unit (CPU); only the cluster-selection network can use a GPU, and it does not require one. The cost per device ranged from about 6~s on the sparsest chip to 28~s on the densest. The four chips order the same way by cost per device and by the number of flakes selected per image, because a crowded chip presents more contours to score and more existing geometry to check every new device against. Alignment and scaling took 45--56\% of each run and was the largest stage in three of the four. Laying out the same four chips by hand, from inspecting the images to drawing every device in mask layout software, occupied the expert for days.

\section{Discussion}\label{sec:discussion}

FlakeForge automates the path from optical images to a lithography-ready layout: it detects and selects usable flakes, aligns and scales every image to chip coordinates and turns each selected flake into a multilayer device layout. Detection tools\cite{uslu2023gmm,uslu2024maskterial,masubuchi2020deeplearning} and stacking pipelines\cite{li2025flakestacking,masubuchi2018autonomous} stop before the last of these steps. The embodied system of Shi et al.\cite{shi2026qumus} does not generate its contact scheme, and it cannot be used without acquiring and setting up its robotic hardware. The layout generator of Lee et al.\cite{lee2026mos2layout} does not generate bond pads, route around neighbouring devices, write an etch layer or handle devices with more than two terminals. FlakeForge instead generates contacts, traces and bond pads for each chip, fits multi-terminal devices to the contour of each selected flake, routes every device around those already placed and writes as many fabrication layers as the user defines (three in the devices demonstrated here). It is open-source software that runs on a standard computer from the optical images a researcher already takes, so it fits into an existing fabrication workflow without new hardware.

The expert comparison shows where the pipeline stands. It produced 119 layouts in 34 minutes, 95 of them usable with at most a minute's correction, where the same task occupied the expert for days. It is less adaptive than the expert, placing 119 devices against 168, and the two selections overlap on 59 flakes. The low overlap does not by itself indicate poor selection: a chip offers more usable flakes than any single layout can use, and the expert judged 24 of the 34 pipeline rejections he examined to be justified. Review during the run narrows the remaining gap with little effort. On Chip A it raised the number of placed devices from 27 to 34, and on Chip C it reduced the number of devices needing any correction from 22 to 9, for about 13 extra minutes. Reviewing during the run is a better use of a user's time than repairing exported geometry afterwards, because each correction is made at the stage that caused the error and everything downstream is recomputed from it.

Several limitations follow from how the pipeline works. Device placement is sequential rather than optimized: candidates are visited in descending score, Hall bars first and LP rings second, and a placement is never revisited, so an early device can block a better arrangement nearby. We also implemented a depth-first branch-and-bound search that optimizes the placement of the whole chip jointly, guided by a heuristic and seeded with the sequential layout. On the chips we tested, it placed one to two more devices on average than the sequential pass, at the cost of several additional minutes of runtime per chip (depending on the search's step limit). Review during the run gains more: on Chip A the search added two devices, whereas the reviewed run added seven. With a user supervising, the search therefore adds little, and the pipeline uses the sequential pass by default.

The usability score is defined by configuration rather than learned, so it encodes one user's judgement. It cannot express where on a flake a defect lies or how a flake compares with its neighbours, since candidates are scored one at a time. The detector scores whole contours and separates flake from substrate by colour, so it cannot use part of a flake, and mono- and few-layer flakes are reachable only when the microscope settings are optimized to give enough contrast between them and the substrate. Both limits could be addressed within the same framework, for example by a second detection pass inside each candidate to find usable sub-regions, or by accepting several colour ranges at once. Where neither suffices, a dedicated detector\cite{uslu2023gmm,uslu2024maskterial,sterbentz2021universalsegmentation} can replace ours through the contour-and-score interface. The device layout stage does not check the flake outlines it receives: if detection draws a single outline around two touching flakes, the device is designed on that combined shape as if it were one flake. Such errors are rare and can be caught at the detection checkpoint by rejecting the outline. Alignment also requires a marker grid and a file-naming convention that encodes position, and the colour reconstruction currently runs in MATLAB (\hyperref[sec:m-reconstruction]{Methods}).

The correction counts also depend on who reviews the layouts. Most corrections were minor, typically a trace that crossed nearby material and would short through it, or a sharp corner, and several reflect the reviewing researcher's own fabrication preferences rather than errors in the layout. The configuration fields that control them, such as minimum gaps, trace geometry and bond pad dimensions, can be set to match other preferences or equipment, so some corrections recorded here would not arise elsewhere. The expert evaluation is one researcher's judgement, made without a pre-agreed rubric, without blinding and without a second rater, and the verdicts on rejected flakes cover only 34 of the 109 flakes in question. Finally, the evaluation is of layouts, not of fabricated devices.

The constructions in FlakeForge (contour-fitted reference rectangles, chip-wide collision-aware routing, per-image alignment and narrow interfaces between stages) are not specific to Hall bars or LP rings. A new device type requires a new geometry definition in the code and its configuration fields, rather than a new pipeline. The marker grid and naming convention used here are simple enough to adopt widely, and chips designed to such a convention would let any automated tool begin from solved alignment and addressing.

\section{Methods}\label{sec:methods}

\subsection{Samples, chips and imaging}\label{sec:m-samples}
4Hb-TaS$_2$ crystals were mechanically exfoliated onto blue tape and transferred to clean Si/SiO$_2$ substrates (250~$\mu$m Si, 285~nm SiO$_2$; 7~$\times$~7~mm) prepatterned with a 5~$\times$~5~mm alignment grid. The grid markers are either Nb (25~nm) or Pt/Cr (20~nm/5~nm) on a 200~$\mu$m pitch. Four chips were used: Chip A , Chip B , Chip C  and Chip D. Images were taken with an HQ Graphene microscope with a 20$\times$ objective. The chip with a different marker design (Fig.~\ref{fig:alignment}) was imaged on a second HQ Graphene microscope with a 20$\times$ objective.

\subsection{Image naming and chip coordinates}\label{sec:m-naming}
Grid cells are named by a two-letter row and column code (AA, AB, \ldots), and each image is named after the cell of its upper left-hand crosshair. A cell name maps to chip coordinates by multiplying its row and column indices by the pitch and adding the grid origin. The two crosshairs in one image are adjacent grid corners, so the chip vector between them is known from the name alone. The pipeline therefore needs no stage coordinates or image metadata, and it does not matter whether the images were acquired manually or with a motorized stage.

\subsection{Flake detection}\label{sec:m-detection}
Each image is smoothed with a Gaussian filter, and pixels outside a configured substrate colour range (in HSV or RGB space) form the candidate mask. Pixels belonging to the chip's own markers are identified by their colour and brightness, dilated and removed from the mask. The mask is cleaned by morphological opening\cite{serra1982}; closing is not applied, because it merges flakes with adjacent contamination. Connected regions are traced as contours\cite{suzuki1985topological} and discarded below a 200-pixel noise floor. Contours outside a configured area range, contours whose bounding box touches the image border (where the flake's extent is unknown), and contours whose solidity falls below a size-dependent threshold are rejected. The solidity threshold decreases linearly from a fixed value for small candidates to a lower configured value for large ones (in our configurations, from 0.72 at 1500 pixels to 0.4--0.5 at 6500 pixels). Surviving candidates are scored on colour fraction, the fraction of contour pixels inside the target colour range, which is also applied as a hard minimum; texture, combining Laplacian variance\cite{pechpacheco2000} and intensity standard deviation over an eroded interior; and isolation, one minus the fraction of a padded neighbourhood covered by non-substrate material in the candidate mask. Their weighted sum is compared with a final threshold. The dark-region alert flags regions below an absolute grey level (40 of 255 in our configurations) within a configured area range and plays no part in selection. All thresholds, weights and colour ranges are configuration fields, set per chip.

\subsection{Colour reconstruction and cluster selection}\label{sec:m-reconstruction}
Each channel is median-filtered (5$\times$5 window), and the image is quantized to eight colours by minimum-variance partitioning of RGB space\cite{wu1991colorquant}. The palette is sorted into a canonical order, and every pixel is assigned to its nearest palette colour without dithering\cite{thomas1991inversecmap}; a second median filter and reassignment then remove isolated pixels. This step uses MATLAB's \texttt{medfilt2} and \texttt{rgb2ind} through the MATLAB Engine for Python; results are cached per image. The cluster-selection network has two residual branches, one over the RGB image and one over the eight binary cluster masks, each downsampled by strided convolutions and fused into one keep/discard logit per cluster (478,648 parameters). The training data came from chips whose layouts had been designed by hand before this work: for each image, an expert had chosen which of the eight clusters to keep, and these choices serve as the labels. This gave 784 image and cluster-map pairs. The network was trained on them with the eight cluster channels randomly permuted per sample so that the network cannot rely on cluster order. One trained checkpoint was used unchanged for every chip in this work. It runs on a CPU and uses a GPU when present.

\subsection{Marker localization}\label{sec:m-markers}
Two masks are built from the reconstruction: a substrate-exclusion mask in HSV space and an Otsu threshold\cite{otsu1979} of the value channel. Two strategies search each mask: normalized cross-correlation\cite{lewis1995} with a template of the whole marker, and detection of the marker's solid core confirmed by evidence of its arms. A third, optional strategy triangulates the marker centre from detected arm segments alone. It exists for crosshairs whose core and one arm are missing from the image itself, a defect specific to some markers on our chips, and it is switched off in the configuration when the markers do not need it, as for the chip with a different marker design (Fig.~\ref{fig:alignment}e), whose markers have no comparable arms. The masks are searched one at a time and pooled only if neither succeeds alone; within a search, candidates from all strategies are scored on a common scale and reduced by non-maximum suppression, and a pair is accepted only if it passes checks on score ambiguity, separation and skew. The accepted cores are refined to sub-pixel position; triangulated positions, which have no core, are kept as found. A reference scale, the median area of small, near-square blobs (overwhelmingly grid dots), sets the size limits for candidate cores and the expected arm spacing; a few limits remain in absolute pixels (core area bounds, suppression radius, minimum marker separation) and are configuration fields. Automatic detection accepts horizontally adjacent pairs; a vertically adjacent pair can be set in the review window.

\subsection{Alignment and scaling}\label{sec:m-alignment}
With marker positions written as complex numbers $p_1, p_2$ in pixels (with the image $y$ axis inverted) and the corresponding chip positions $r_1, r_2$ in micrometres, the similarity transform $r = r_1 + s\,(p - p_1)$ with $s = (r_2 - r_1)/(p_2 - p_1)$ gives scale $|s|$ and rotation $\arg s$ in one step and admits any axis-aligned marker adjacency.

\subsection{Polygon extraction and export}\label{sec:m-polygons}
The reconstruction is smoothed, edges are found with the Canny detector\cite{canny1986}, small gaps are closed morphologically, and enclosed regions are filled and traced as polygons, each assigned to the layer of its colour cluster; polygons above a vertex limit are split. Polygons are transformed to chip coordinates and written in CIF into a working copy of the chip's master layout, which itself is never modified.

\subsection{Device geometry}\label{sec:m-geometry}
The device orientation is that of the contour's minimum-area bounding rectangle, found by rotating calipers\cite{toussaint1983}, and all geometry is computed in that frame. For Hall bars, the contour is rasterized on a 0.1~$\mu$m grid in that frame, eroded by one cell so that the result cannot cross the true boundary, and searched for the largest axis-aligned inscribed rectangle with the row-histogram and monotonic-stack method for maximal rectangles in a binary matrix, a relative of the largest-empty-rectangle problem\cite{chazelle1986}; the result is exact for the chosen orientation at the raster resolution. A flake is excluded from Hall placement if this rectangle is shorter or narrower than configured minima or its length-to-width ratio is below a configured value (2 in our configurations). The rectangle is shrunk by configured margins to an inner rectangle and grown to a contact area. The etch mask removes the flake outside a bar formed by the inner rectangle and six voltage probes extending from its long sides to the edge of the reference rectangle, so the channel and probes have their configured dimensions regardless of the shape of the flake. Two current contacts span the full width at either end; six voltage contacts sit three per long side at the quarter, half and three-quarter positions of the inner rectangle, and placement is abandoned if adjacent contacts cannot keep their minimum separation. Each voltage contact keeps its configured width inside the inner rectangle and widens linearly to a configured multiple of that width (1.5 here) at its outer end; the four outer contacts bend by a configured angle (30$^\circ$ here) away from the centre pair. For LP rings, the reference is the minimum enclosing rectangle. The etch mask is built at its centre from a solid inner square, which opens the hole, and two C-shaped cuts that isolate the loop from the surrounding flake, each interrupted to leave one lead; hole size, wall width, lead width and cut separation are configuration fields, with the hole and wall chosen from configured ranges. Two voltage and two current contacts are placed symmetrically along the length axis; each voltage contact widens only on the side away from the etch mask, keeping its configured clearance from it. Every dimension in this section is a configuration field.

\subsection{Routing and chip-level placement}\label{sec:m-routing}
Each device reports its contact endpoints as position, outward direction and width. A shared routine places the required bond pads inside a square work area around the device without overlap, keeping a minimum spacing to the bond pads of earlier devices, and connects each contact to its bond pad with a single straight, tapered trace. Placed geometry, together with the bodies of the other selected flakes, forms a chip-wide occupied region; a device whose geometry intersects this region or crosses the chip-border margin is discarded entirely. Candidates are visited in descending usability score across all images of the chip, first for Hall bars up to a requested number and then for LP rings on the remaining flakes up to a second requested number. Fabrication layers are defined in the configuration; the configuration used here writes three, matching the layer convention of the chip's master layout: fine contacts, coarse traces and bond pads, and the etch mask.

We also implemented an alternative to the sequential pass: a depth-first branch-and-bound search over placements for the whole chip. Flakes are taken in descending score, and for each the search either skips the flake, places a Hall bar, or places an LP ring with its bond pads on one of two sides at one of ten pad distances, each placement computed given the devices already placed. Combinations are enumerated in a fixed flake order so that each set of devices is generated only once. The objective is the number of devices placed, with a Hall bar weighted 1.5 times an LP ring; a branch is pruned when its value plus the best value each remaining flake could reach alone on an empty chip cannot exceed the best layout found so far. The search starts from the sequential layout, so it never returns fewer devices, and stops after a fixed number of steps. Its gain was small compared with the added runtime (Discussion), and the results reported here use the sequential pass.

\subsection{Review interfaces}\label{sec:m-review}
Both review windows are optional and are enabled per stage in the configuration. Marker corrections are saved to disk and reused by all later runs until they are cleared; flake selections made in the detection review apply to the current run. A form for interactive device-layout parameters exists in the code but is not part of the pipeline runs reported here.

\subsection{Domain-expert evaluation}\label{sec:m-expert}
One domain expert, working from the same images, independently selected every flake they would have built a device on, aiming at the densest usable layout, and separately inspected every device the pipeline placed, recording no correction, small fix (up to one minute spent per fix), big fix (reshaping some or all contacts) or deletion. The expert knew which layouts came from the pipeline, no rubric was agreed in advance, and no second rater was used, so no inter-rater agreement is reported. Verdicts on pipeline rejections were recorded opportunistically for 34 of the 109 flakes the expert selected and the pipeline did not use, so they bound rather than estimate the rejection error rate.

\subsection{Runtime measurement}\label{sec:m-runtime}
Each run records the wall-clock time of every stage and of the whole run. The unattended runs used a laptop with an Intel Core i7-12700H, 32~GB of memory and an NVIDIA RTX 4060 Laptop GPU; the reviewed Chips A and C runs used a desktop with an Intel Core i7-14700, 32~GB of memory and no GPU.

\backmatter

\phantomsection
\addcontentsline{toc}{section}{Data availability}
\section*{Data availability}
A toy dataset of optical images for running the pipeline is available at \url{https://drive.google.com/drive/folders/1PFAKjvdyGy75H3g4cxbodw90LaiyOWCD?usp=drive_link}.

\phantomsection
\addcontentsline{toc}{section}{Code availability}
\section*{Code availability}
FlakeForge is available at \url{https://github.com/ARPESTechnion/chip_design.git}. The pipeline is written in Python; the colour reconstruction requires MATLAB with the Image Processing Toolbox. The trained cluster-selection model is available, together with the toy dataset, at \url{https://drive.google.com/drive/folders/1PFAKjvdyGy75H3g4cxbodw90LaiyOWCD?usp=drive_link}.

\phantomsection
\addcontentsline{toc}{section}{Acknowledgements}
\section*{Acknowledgements}
We thank Amit Kanigel for funding this work and for providing the samples and substrates. We thank Irena Feldman for growing the crystals, and Michael Shlafman for writing the original MATLAB code for alignment and scaling and for the initial design of the alignment-marker grid. We also thank Amit Keren, Ido Schwartz and Gad Koren for use of their equipment.

\phantomsection
\addcontentsline{toc}{section}{Author contributions}
\section*{Author contributions}
O.B. and I.M. contributed equally to this work. O.B. developed the software. I.M. exfoliated the samples, designed the device layouts and performed the expert evaluation. Both authors wrote the manuscript.

\phantomsection
\addcontentsline{toc}{section}{Competing interests}
\section*{Competing interests}
The authors declare no competing interests.

\bibliography{references}

\end{document}